# Network formation and efficiency in linear-quadratic games: An experimental study

Gergely Horvath[1]

*Division of Social Sciences, Duke Kunshan University*

**Abstract**

*We experimentally study effort provision and network formation in the linear-quadratic game characterized by positive externality and complementarity of effort choices among network neighbors. We compare experimental outcomes to the equilibrium and efficient allocations and study the impact of group size and linking costs. We find that individuals overprovide effort relative to the equilibrium level on the network they form. However, their payoffs are lower than the equilibrium payoffs because they create fewer links than it is optimal which limits the beneficial spillovers of effort provision. Reducing the linking costs does not significantly increase the connectedness of the network and the welfare loss is higher in larger groups. Individuals connect to the highest effort providers in the group and ignore links to relative low effort providers, even if those links would be beneficial to form. This explains the lack of links in the network.*

---

[1] Division of Social Sciences, Duke Kunshan University. Address: Division of Social Sciences, Duke Kunshan University, No. 8 Duke Avenue, Kunshan, Jiangsu Province, China 215316. E-mail address: horvathgergely@gmail.com. Tel: +86 155-01683821.

This research was supported by the National Natural Science Foundation of China (NSFC), under the 2021 NSFC Research Fund for International Excellent Young Scientists (RFIS-II) program, grant nr. 72150610501. The experiment was approved by the Institutional Review Board of the Duke Kunshan University on 23/03/2022, approval nr.: 2022GH014. The experiment was registered in the AEA RCT Registry with ID AEARCTR-0009999.

## 1. Introduction

In many important applications, collaboration takes place in a social network, in which individuals or firms collaborate with their direct neighbors in the network and their collaborative efforts have positive spillovers for more distant agents in the network. These applications include co-authorship among scientists (Ductor et al., 2014; Essers et al., 2022; Hsieh et al., 2018; Newman, 2004), R&D collaboration among firms (Dawid and Hellmann, 2014; Goyal and Moraga-Gonzalez, 2001; König et al., 2012), educational attainment (Calvo-Armengol et al., 2009), criminal activity (Lee et al., 2021; Lindquist and Zenou, 2019), among others. Agents in these strategic situations typically choose a costly effort level which benefits themselves and their direct neighbors and, in the longer run, they also form new and delete old links changing the set of their direct neighbors. In addition, complementarities exist between the effort choices of neighbors, as one's optimal effort level rises if their neighbors increase their effort levels. Assuming a linear-quadratic utility function that captures these features of the interactions, the game theoretic network literature has derived two important results. Firstly, the Nash equilibrium effort level of an individual is determined by their network position as captured by the Katz-Bonacich centrality[2] (Ballester et al., 2010). Secondly, when individuals make linking decisions strategically, the equilibrium network has a core-periphery structure in which core members connect to each other and the periphery, while peripheral nodes connect only to the core (Hiller, 2017). This type of network structure has been widely observed in empirical applications (Fricke and Lux, 2015; Juhász et al., 2020; König et al. 2012; Leydersdorff and Wagner, 2008), supporting the predictions of the model.

An economically relevant question is how far individuals can exploit the benefits of collaboration in these situations. Both choosing a higher effort level and creating a link by a given individual impose positive externalities on others who are or become directly connected to them. This implies that the equilibrium effort choices and network structure are typically not socially efficient, meaning that not all benefits are exploited (Ballester et al. 2010, Hiller, 2017). However, Gallo and Yan (2021) show in an experiment on *fixed networks* that in smaller groups, individuals can coordinate on a significantly higher effort than the Nash equilibrium level, which increases their payoffs. The authors refer to this phenomenon as the emergence of a *collaborative social norm*. It is an open question, however, if the overprovision of effort would emerge in an *endogenously formed network* as well or if individuals would form the efficient network structure, which also takes the core-periphery structure, but may be different from the equilibrium network (Hiller, 2017).

---

[2] The Katz-Bonacich centrality is the total number of direct and indirect neighbors of an individual where neighbors are weighted by distance to the individual.

Motivated by these questions, in this paper, we present the results of an online experiment where we study network formation and effort provision in the described setting of positive externality and complementarity embedded in the linear-quadratic utility function. We analyze if individuals will form the Nash equilibrium network and provide the Nash equilibrium effort level and study equilibrium selection in treatments with multiple equilibria. We also analyze the efficiency of the outcomes focusing on whether individuals can coordinate on an outcome with higher effort and payoffs than the Nash equilibrium levels, or alternatively, there are some other deviations from the equilibrium that reduce welfare. In addition, we study how the answers to these questions depend on parameters like the group size and the cost of establishing a link.

In the experiment, individuals make decisions for 30 periods in fixed groups. In each period, they choose a costly effort level and the other group members they want to initiate a link with. Network links are formed unilaterally and cost only for the party who initiated the link. Individuals benefit from their own effort level and the effort level of their direct neighbors in the network, independently of who initiated the link. Payoffs are determined by the linear-quadratic utility function based on the network formed in a given period. Participants are paid after the sum of payoffs received over the 30 periods of the game.

We introduce five treatments in a 2x2 factorial design with an additional treatment. On the one hand, we vary the group size between a smaller group of five individuals and a larger group of nine individuals. We conjecture that under the larger group size, it is harder to coordinate on a single equilibrium when there are many and it is harder to coordinate on an outcome with effort provision above the equilibrium level. On the other hand, we vary the linking costs which moderate the number of equilibria. For low linking costs, the only equilibrium is the complete network. For high linking costs, there are three equilibria, the empty, complete and star networks.[3] We conjecture that under multiple equilibria it is harder for the participants to establish an equilibrium. To generate the exact same equilibria for the smaller and larger groups in the described way, we also need to adjust other parameters of the payoff function together with the group size, which does not allow us to compare those treatments and study the impact of group size in a clean way. For this reason, we introduce a fifth treatment where we hold the payoff parameters fixed, and compare the smaller and larger groups for the low linking cost case in which the only equilibrium is the complete network. We note that the efficient network is the complete one in all five treatments.

Our main results based on the last 10 periods of the experiment can be summarized as follows. Regarding the network structures formed, we find that in the treatments with larger group size, individuals never form an equilibrium network, independently of the number of equilibria. In the treatments with smaller group size, they form the complete network in 25.3% of the cases when linking costs are low, and in 18.7% of the cases when linking costs are high. These fractions go up to 76.7% and 66.7%, respectively, when we allow for mistakes being made. The other equilibrium networks for low linking costs, the empty and star networks, are never played. The complete network is thus the most focal outcome, but is not played in the overwhelming majority of the cases, even if it would create the highest payoffs for the participants.

---

[3] In the empty network there are no links, while in the complete network all links are present. These structures are special cases of the core-periphery network structure, in the empty network all nodes are in "periphery", while in the complete network all nodes are in the "core". In the star network, a single node connected to everyone else is in the core while all other nodes are in the periphery and connected only to the core.

Considering all treatments, we find that networks are under-connected on average relative to the most focal equilibrium of complete network. We obtain that only about 60-80% of all potential links are present in the network and decreasing the linking costs does not significantly increase the connectedness of the network. We also find substantial individual variation in the number of links created. For example, in the treatments with larger group size, the difference between the most and least connected individuals is about 4 links on average.

Turning to the effort provision, our main finding is that individuals provide significantly higher effort than the Nash equilibrium effort level *on the network they created*. This holds for four out of our five treatments. We thus find overprovision of effort in the endogenous network, similarly to the results obtained in Gallo and Yan (2021) for fixed networks. We use a modified version of the myopic best-response dynamics model to rationalize effort choices and explain treatment differences in the overprovision of effort. Despite the described higher effort provision, individuals earn significantly lower payoffs on average compared to what they could earn in the most focal Nash equilibrium of complete network, which also implies that the welfare is lower than in the efficient allocation. We find that only about 65% of potential equilibrium payoffs are realized in the treatments with smaller group size, which goes down to 33.6% in the comparable treatment with larger group size. The welfare loss is significantly higher in the larger than in the smaller group. The main reason for the welfare loss is the low connectivity of the network, which limits the welfare-improving positive spillover effects among the individuals. We indeed verify that creating the vast majority of the missing links would be beneficial for the individuals and the benefits missed are non-negligible.

To understand the sources of welfare loss better, we study the factors that influence the linking decisions. Individuals are more likely to form a link if the benefits of it are higher. However, they associate the benefits of a link with the effort provided by the neighbor at the end of the link, ignoring other important factors that affect the benefits (such as their own effort and the linking costs). To understand if this can explain the missing links, we run an additional experimental treatment in which we provide exact information on the benefits of each link on the feedback screen at the end of each period. We show that the number of links formed in the group are not statistically significantly different between this additional treatment and the original setting where no information on link benefits is provided. The misunderstanding of linking benefits thus cannot explain the lack of links. Further analyzing the linking decisions of individuals, we show that they tend to form links based on the relative ranking of effort provision in the group: they form links to top effort providers in the group and ignore links to others who are ranked low relative to other group members, despite these links would be beneficial to form too. This explains why links are missing from the network.

In sum, we find that individuals do not exploit all welfare gains from collaboration within groups when the network is endogenous. Despite they overprovide effort relative to the equilibrium level, their payoffs are substantially lower than the equilibrium payoffs. This is because they do not create all links that would allow them to benefit from others' effort and create further positive spillovers of effort provision. We obtain that the welfare loss is higher in larger groups and that smaller linking costs do not induce individuals to create more links. The lack of links can be explained by the tendency of individuals to connect to top effort providers in the group and ignore otherwise beneficial links to relative low effort providers. Our results suggests that networked systems of collaboration tend to be under-connected and there is a need for policy interventions to encourage link formation which would substantially increase welfare.

Our paper contributes to the literature on network game experiments which was recently reviewed in Choi et al. (2016). One part of this literature studied games played on fixed networks. This includes prisoner's dilemma games (Rand et al., 2014), coordination games (Berninghaus et al., 2002; Cassar, 2007), local public goods games (Caria and Fafchamps, 2019; Fatas et al., 2010; Rosenkrantz and Weitzel, 2012; Suri and Watts, 2011), games of binary actions with strategic complements and substitutes (Charness et al., 2014), and status competition games (Antinyan et al., 2020). The most related paper is Gallo and Yan (2021) who study effort provision in the game with linear-quadratic utility function on fixed networks. As mentioned earlier, they show that individuals provide higher effort than the Nash equilibrium level by coordinating on a cooperative social norm when groups are relatively small. We study whether similar findings hold on endogenous networks as well[4] and show that despite overprovision of effort, there are significant welfare losses in these networks due to the lack of links formed.

Another part of the experimental network game literature studies network formation. Several earlier papers (Berninghaus et al., 2007; Callander and Plot, 2005; Caria and Fafchamps, 2020; Falk and Kosfeld, 2012; Goeree et al., 2009) test the predictions of the network formation model in Bala and Goyal (2000), in which benefits depend on the number of direct and indirect connections of the individual and not on effort choices. The most related paper is Caria and Fafchamps (2020) who find welfare loss in network formation due to individuals making suboptimal linking decisions based on degree instead of myopic best response. We also find welfare loss due to suboptimal link formation but the mechanism for the result and the context are different. More recent papers (Choi et al., 2022; Rong and Houser, 2015; van Leeuwen et al., 2020) focus on the laboratory validation of the local public goods game in Galeotti and Goyal (2010), which exhibits strategic substitution instead of complementarity. Network formation has been also studied in the prisoner's dilemma and minimum effort games, where it is shown that endogenous linking decision supports cooperation (Gallo and Yan, 2015; Rand et al., 2011) and coordination on the efficient equilibrium (Riedl et al., 2016), respectively.

Our contribution to this literature is that we consider network formation in the case of positive externality and complementarity in effort choices as embedded in the linear-quadratic utility function. This model has been theoretically studied on fixed networks in Ballester et al. (2010) and Bramoulle et al. (2014), and in endogenous networks in Hiller (2017) and Baetz (2015). However, there is no previous experimental paper studying network formation, equilibrium selection and efficiency in this game. Our paper aims to fill this gap.

---

[4] We note that it is not evident that the same results hold on fixed and endogenous networks. Vega-Redondo (2016) reviews the network game literature and shows that results obtained on fixed networks typically change when considering endogenous networks.

The paper is organized as follows. Section 2 introduces the theoretical background and the research questions. In Section 3, we describe the experimental setting and treatments. In Section 4, we discuss the main results of the paper and in Section 5, we study behavior at the individual level. Section 6 concludes the paper.

## 2. Theoretical framework and research questions

In this section, we introduce the theoretical framework of the study and state the research questions we aim to answer. Our framework follows the one-sided network formation model introduced in Hiller (2013, 2017).

Suppose a society of $N$ individuals who are connected by a network. A network is given by the $N \times N$ adjacency matrix $G$ in which $g_{ij} = 1$ if there is a link connecting individual $i$ to individual $j$, and $g_{ij} = 0$ otherwise. The links are mutual: if individual $i$ is connected to individual $j$, it is also true that individual $j$ is connected to individual $i$. This means that the matrix $G$ is symmetric (if $g_{ij} = 1$ then $g_{ji} = 1\ for\ \forall ij$). The set of direct neighbors of individual $i$ in network $G$ is denoted by $N_i(G) = \{j \in N: g_{ij} = 1\}$, the number of neighbors is the cardinality of this set, denoted by $\eta_i = |N_i(G)|$. Using the network analysis jargon, we refer to the number of neighbors as the degree of the node.

In the model, each agent $i$ makes two decisions in a *simultaneous way*. On the one hand, agent $i$ chooses a positive effort level $x_i \in X = [0, +\infty)$. On the other hand, she can initiate network links to others in the group. Formally, she chooses $g'_{ij} \in \{0,1\}$ for each $j \in N\backslash i$, where $g'_{ij} = 1$ means that individual $i$ initiates a link with individual $j$. We assume that individuals can *establish links unilaterally*. Therefore, a link exists in the network if at least one of the agents involved in the link chooses to initiate it: $g_{ij} = 1$, if either $g'_{ij} = 1$, or $g'_{ji} = 1$. We focus on the one-sided link formation case because it presents a simpler strategic situation for the participants of the experiment than the two-sided link formation in which mutual consent is required to establish a link. As shown in Hiller (2017), the general structure of the equilibrium networks does not depend on whether links are established unilaterally or by mutual consent.[5] The linking choices of all individuals will establish the adjacency matrix $G$. Further, we introduce the following notation. The set of strategies for the linking choice of an agent is $G' = \{0,1\}^{N-1}$ and the set of strategies is $S = X \times G'$. The vector of all effort choices of the $N$ individuals is denoted by $x = (x_1, x_2, \ldots, x_N)$.

The payoffs resulting from these choices are given by the following equation:

$$\pi_i(x, G) = \theta x_i - \frac{\beta}{2} x_i^2 + \lambda x_i \sum_{k \in N_i(G)} x_k - \kappa \eta_i' \qquad (1)$$

where $\theta > 0, \beta > 0$ and $\lambda > 0$. This payoff function is the so-called linear-quadratic utility function which is widely studied in the theoretical and empirical literature on social networks (see e.g., Ballester et al., 2010; Calvó-Armengol et al., 2009; König et al., 2014). According to the first term of the function, individual $i$ benefits from her own effort, but effort is costly, the second term represents the quadratic cost function of effort. The third term shows that individual $i$ benefits from the effort of their direct neighbors in the network. Note that this benefit is independent of who has initiated the link, agent $i$ or her neighbors. This formulation of benefitting from the neighbors' effort leads to complementarity between the effort choices of direct neighbors: when the neighbors exert a higher total effort, individual $i$ has more incentives to exert effort as well.[6] Finally, the last term of the payoff function $\kappa \eta_i'$ captures that individuals pay for the links that *they initiated*. $\kappa$ represents the cost of initiating a link and $\eta_i'$ stands for the number of links initiated by individual $i$, which is given by $\eta_i' = |\{j \in N : g_{ij}' = 1\}|$.

---

[5] We note that equilibrium effort provision does not depend on whether the links are formed by unilateral or mutual consent. In addition, the structure of equilibrium networks is similar: for low linking costs, the unique equilibrium is the complete network; for high linking costs, it is the empty network; and for intermediate values of the linking cost parameter, the equilibrium networks are nested split graphs. However, the cost cutoffs are not be the same for the cases of unilateral and mutual link formation.

[6] The best-response function of effort choice is given by: $x_i^* = \frac{\theta + \lambda \sum_{k \in N_i(G)} x_k}{\beta}$, where the optimal effort of agent $i$ positively depends on the sum of effort levels of their neighbors.

The Nash equilibrium (NE) of this game is defined as follows.

**Definition 1.** A strategy profile $s^* = (x^*, G^*)$ is a Nash equilibrium if and only if for any $i \in N$ and every $s_i \in S$, $\pi_i(s^*) \geq \pi_i(s_i, s^*_{-i})$.

Hiller (2017) shows that the Nash equilibrium network configuration is a nested-split graph. A nested-split graph is defined as follows:

**Definition 2.** A network G is a nested-split graph if and only if $[g_{il} = 1 \text{ and } \eta_k(G) \geq \eta_l(G)] \Rightarrow g_{ik} = 1$.

This means that, in a nested-split graph if individual $i$ is connected to individual $l$, and individual $k$ has at least as many neighbors as individual $l$, then it must be that individual $i$ is also connected to individual $k$. In other words, in a nested-split graph there is a strict hierarchy among the nodes, in which a node with a higher number of neighbors is connected to all nodes with a lower number of neighbors. We note that the empty network, in which there are no links, and the complete network, in which all possible links are present, are also nested-split graphs. In addition, the star network is a nested-split graph too. In a star network, the center of the star connects to all others in the group, while others connect only to the center of the star. We also note that nested-split graphs are a subset of the empirically widely observed core-periphery networks (Hiller, 2017; König et al., 2014), in which core members occupy the connected center of the network, while peripheral nodes are connected only to the core, but not to each other.[7]

The previous game-theoretic literature on social networks has derived the following results regarding the game defined above:

    ***Result 1.*** On any given network, the vector of NE effort levels is determined by the Katz-Bonacich centrality (Ballester et al., 2010): $x^*(G) = \frac{\theta}{\beta} b(G, \lambda)$ where $b(G, \lambda) = [I -$

---

[7] Formally, in a core-periphery network nodes can be partitioned into two disjoint sets: the core $C(G)$ and the periphery $P(G)$. It holds that $g_{ij} = 1$ for all $i, j \in C(G)$ and $g_{ij} = 0$ for all $i, j \in P(G)$.

$\frac{\lambda}{\beta}G]^{-1} * \mathbf{1}$ is the Katz-Bonacich centrality. Here $I$ is the N-dimensional identity matrix, and $\mathbf{1}$ is the N-dimensional vector of ones. The Katz-Bonacich centrality of individual i is determined by the total number of direct and indirect neighbors reachable from i's position by following the links of the network. This sum is weighted by the distance in such a way that individuals further away from individual i in the network receive smaller weights.

***Result 2.*** In general, the effort choice is not efficient since there are positive externalities in the effort choices of individuals (Ballester et al., 2010; Gallo and Yan, 2021). The efficient effort level is higher than the equilibrium level.

***Result 3.*** In any Nash equilibrium, the network is a nested-split graph (Hiller, 2017).

***Result 4.*** There exists two cutoffs of the linking cost $\kappa_1$ and $\kappa_2$ such that for linking costs smaller than $\kappa_1$ ($\kappa < \kappa_1$), the unique NE network is the complete graph; for linking costs larger than $\kappa_2$ ($\kappa > \kappa_2$), the unique NE network is the empty network. For intermediate linking cost values ($\kappa \in [\kappa_1, \kappa_2]$), there are multiple equilibria: the complete and empty networks are NE, possibly alongside with other nested-split networks (Hiller, 2013). Note that the thresholds $\kappa_1$ and $\kappa_2$ depend on the group size $N$ and the parameters of the payoff function (1).

***Result 5.*** The efficient network structure is either the empty or the complete network (Hiller, 2017).

We note these theoretical predictions are based on the equilibrium of the static game, while in the repeated setting of the experiment, there might be some dynamic strategic considerations as well. For example, a dynamic grim-trigger strategy can be to provide high effort until others provide high effort which will increase payoffs for everyone. High effort will then make link formation more beneficial and lead to the formation of a complete network. If someone deviates and provides relatively low effort, others would decrease effort provision as well and withdraw their links to the person. This would effectively be a payoff-reducing punishment for the deviating individual. Based on these considerations, the complete network with high effort, even higher than the Nash equilibrium level, can be an outcome in the repeated game. We will examine if the experimental outcomes can be explained by these dynamic effects.

Based on this theoretical background, we ask the following research questions:

*RQ1. Will individuals form an equilibrium network and choose the equilibrium effort level?*
*RQ1a. If an equilibrium is played, which equilibrium will be selected in case of multiple equilibria?*
*RQ1b. If no equilibrium is played, what will be the properties of the network structure formed and the effort level chosen?*

Nash equilibria constitute a natural benchmark of behavior that maximizes individual payoffs given the strategies chosen by others. However, the game is relatively complex as individuals choose both their partners and an effort level, each decision involving trade-offs. This complexity may make it difficult for them to figure out the optimal strategy. Moreover, in the case of multiple equilibria, individuals have to coordinate on one of them. We conjecture that the complete network, which is always an equilibrium when there are multiple equilibria, will be the most focal outcome. This is because it involves the simplest strategy of connecting to everyone else in the group, without making a distinction among the other group members.

*RQ2. Will individuals be able to coordinate on a more efficient outcome than the one associated with the equilibrium?*

Regarding the effort choice, Gallo and Yan (2021) shows experimentally that on *fixed networks* of 15 and 21 individuals, experimental participants choose the NE effort level. In smaller networks of 9 individuals, however, experimental participants establish a collaborative social norm in which they provide significantly higher effort levels than the NE level and this increases their payoffs. Individuals are thus able to internalize the positive externalities of the effort choice to some extent. We ask if individuals will overprovide effort in a situation when individuals not only choose their effort levels but also their neighbors in the network, which may be deemed as more realistic in the longer run.

*RQ3. How do the answers to RQ1 – RQ2 depend on the group size $N$ and the cost of linking $\kappa$?*

The answers to RQ1 – RQ2 may depend on the size of the group as in a larger group it may be more difficult to coordinate on a particular equilibrium when there are multiple equilibria. As Gallo and Yan (2021) show group size also influences whether individuals overprovide effort or follow the NE effort level. Regarding the linking cost $\kappa$, it moderates the number of equilibria (see Result 4 above). It may be easier to play a NE and overprovide effort if there is only one equilibrium than in case of multiple equilibria. We consider different values of $\kappa$ such that in some treatments there is a single equilibrium, while in others there are three equilibria.

## 3. Experimental design

### 3.1 General experimental framework

The experiment follows the structure of the theoretical model described above. We randomly form groups of $N$ participants, the groups remain the same along the multiple periods of the experiment. The experiment is anonymous, participants are referred to by ID numbers that do not change during the experiment. In each period, participants decide about whom to establish a network link with by clicking on boxes placed next to the ID numbers of the other group members (a screenshot is shown in Figure A1 in the Appendix). Simultaneously, individuals choose an effort level between 0 and 20 where decimal numbers are allowed.[8] All individuals make these decisions simultaneously without knowing the decisions made by others. They had a 1-minute time limit to submit their choices. Once everyone has made their decisions, the network is generated based on the choices of the participants and payoffs are computed according to equation (1). Participants receive detailed feedback on the outcomes of the period, this includes their effort level chosen; the list and number of group members they initiated links to; the list and number of group members who became their network neighbors; their payoffs in total and broken down to costs and benefits; the effort choices and list of network neighbors of other group members; and a graph showing the established social network.[9]

---

[8] We did not choose a higher upper limit for the effort choice because it would imply very high costs for participants due to the quadratic costs of effort and result in negative payoffs, which should be avoided in experiments.

The game is played for 30 periods. The payoff of the participant is the sum of per period payoffs earned over the 30 periods. Note that payoffs can be negative in a given period, however, if the sum of payoffs over all 30 periods is negative, the participant receives zero payoffs from the network game. The 30 periods of network game is preceded by 5 practice periods and followed by a questionnaire as explained in detail below.

### 3.2 Treatments

We introduce five treatments to this design in a 2x2+1 structure where we vary the group size and the cost of linking. Table 1 summarizes the experimental treatments. The following parameter values of the payoff function are constant across all treatments: $\theta = 10, \beta = 4$. Other parameter values are varied based on the implied equilibrium network configurations as shown in the Table 1.

**Table 1: Experimental treatments, parameter values and equilibrium networks**

|  | **Group size: $N = 5$** | **Group size: $N = 9$** | **Group size: $N = 9$** |
|---|---|---|---|
| **Low cost of linking** | **Treatment:** *N5_LowCost* <br> **Parameter values** <br> $\lambda = 0.4$ <br> $\kappa = 1$ <br> **Equilibrium networks** <br> • Complete | **Treatment:** *N9_LowCost1* <br> **Parameter values** <br> $\lambda = 0.25$ <br> $\kappa = 1$ <br> **Equilibrium networks** <br> • Complete | **Treatment:** *N9_LowCost2* <br> **Parameter values** <br> $\lambda = 0.4$ <br> $\kappa = 1$ <br> **Equilibrium networks** <br> • Complete |
| **High cost of linking** | **Treatment:** *N5_HighCost* <br> **Parameter values** <br> $\lambda = 0.4$ <br> $\kappa = 3.9$ <br> **Equilibrium networks** <br> • Empty <br> • Star <br> • Complete | **Treatment:** *N9_HighCost* <br> **Parameter values** <br> $\lambda = 0.25$ <br> $\kappa = 2.5$ <br> **Equilibrium networks** <br> • Empty <br> • Star <br> • Complete | |

---

[9] A screenshot of the feedback screen is available in Figure A2 in the Appendix.

We vary the group size between two values $N = 5$ and $N = 9$. For each group size, we consider two levels of linking costs: 1) for low linking costs, the only equilibrium is the complete network; 2) for high linking costs, there are three equilibria. These three equilibria are the empty, complete and star networks. To achieve these fixed equilibrium network structures, however, we also need to adjust the parameter value $\lambda$ between treatments of different group sizes $N$. This is because the linking cost thresholds that moderate the set of equilibria depend on $N$ and $\lambda$ (as explained in *Result 4* in the previous section). More specifically, we set $\lambda = 0.4$ for the treatments with $N = 5$ and $\lambda = 0.25$ for the treatments with $N = 9$. To avoid $\lambda$ being a confounding factor of group size, we introduce the treatment *N9_LowCost2* where we set $\lambda = 0.4$ for $N = 9$ as well, implying the unique equilibrium of complete network. This allows us to make a clean comparison between the two group sizes by comparing *N5_LowCost to N9_LowCost2.*

For each treatment, we compare the experimental outcome to the NE predictions (answering RQ1) and analyze whether individuals can establish a collaborative social norm that increases effort levels above the Nash equilibrium value (answering RQ2). In *N5_HighCost* and *N9_HighCost*, we can study equilibrium selection among the three equilibrium networks for each group size and answer RQ1a. In addition, our treatments allow us to make the following comparisons to answer RQ3. Comparing *N5_LowCost* to *N5_HighCost*, as well as, *N9_LowCost1* to *N9_HighCost*, we are able to study the impact of linking costs for a given group size $N$. Comparing *N5_LowCost* to *N9_LowCost2*, we can analyze the impact of group size for holding all other parameters and the equilibrium network structure fixed.

Table 2 shows the predicted effort levels and payoffs in the Nash equilibria. We can see that the complete network gives the highest payoffs when there are multiple equilibria and thus is the Pareto-dominant equilibrium. We also compute for each equilibrium network, the effort levels that would ensure the highest possible total payoffs in the group. These are higher than the equilibrium effort levels due to the positive externalities. We can see that in all treatments, the complete network gives the highest total payoffs and it is the efficient network.

## 3.3 Procedures

The experiment was conducted online between July and September 2022 with participants recruited via the Prolific website (Palan and Schitter, 2018). There is broad evidence that online experiments are reliable and produce comparable results to lab experiments (Amir and Rand, 2012; Kroher and Wolbring, 2015; Paolacci et al., 2010; Rand, 2012; Suri and Watts, 2011) and they have been applied in network experiments as well (see e.g., Gallo and Yan, 2015, 2021; Rand et al., 2011). The experiment was programmed in o-Tree (Chen et al., 2016). The experimental instructions are attached in the Appendix and contain screenshots about the decision and feedback screens.

The flow of the experiment was as follows. Once the experiment was posted on Prolific, participants could enroll in the experiment. They started by reading the instructions, after which they answered a set of control questions. Participants could not go further with the experiment until they answered all question correctly. This was followed by five practice periods in which they played the network game against the computer which allowed them to get used to the experimental interface and understand the trade-offs involved in the decisions they had to make.[10] The practice periods were not incentivized. After finishing the practice periods, groups were formed in the order of arrival of participants to this stage[11] (a standard procedure described in Arechar et al., 2018).[12] Once the group was formed, participants played the 30 periods of the network formation game as described above.

**Table 2: Equilibrium and efficient effort levels in the experimental treatments**

| Treatment | N5_LowCost | N5_HighCost | N9_LowCost1 | N9_LowCost2 | N9_HighCost |
|---|---|---|---|---|---|
| **Nash equilibrium** | | | | | |

---

[10] The computer played the same randomly chosen strategy for all participants and in all practice periods. Fixing the opponents' strategy in these practice periods was meant to support the participants' learning about the payoff function in a simplified environment. Nevertheless, participants made the same effort and linking choices, and received the same feedback as in the experiment itself.

[11] In the Appendix, we provide robustness checks of our results with respect to the time to form a group. We compare behavior in groups that are formed relatively fast to those that are formed relatively slowly. We find no significant difference between these groups.

[12] Participants who could not be matched with the necessary number of other participants to form a group, exited the study and received the show-up fee.

| | | | | | |
|---|---|---|---|---|---|
| **Effort** | Complete: 4.17 | Empty: 2.5<br>Star: {3.65, 2.86}<br>Complete: 4.17 | Complete: 5 | Complete: 12.5 | Empty: 2.5<br>Star: {3.87, 2.74}<br>Complete: 5 |
| **Payoffs** | Complete: 32.72 | Empty: 12.5<br>Star: {26.58, 12.51}<br>Complete: 26.92 | Complete: 46 | Complete: 308.5 | Empty: 12.5<br>Star: {29.97, 12.54}<br>Complete: 40 |
| **Efficient allocation** | | | | | |
| **Efficient effort** | Complete: 12.5 | Empty: 2.5<br>Star: {5.36, 3.57}<br>Complete: 12.5 | Complete: 20 | Complete: 20 | Empty: 2.5<br>Star: {5.7, 3.2}<br>Complete: 20 |
| **Efficient payoffs** | Complete: 60.5 | Empty: 12.5<br>Star: {26.82, 13.95}<br>Complete: 54.69 | Complete: 195.80 | Complete: 674.84 | Empty: 12.5<br>Star: {28.55, 13.57}<br>Complete: 189.8 |

At the end of the experiment, participants filled out a survey that included questions about demographic information (gender, highest education, nationality, age, student and work statuses), an incentivized cognitive-reflection test (Frederick, 2005), an incentivized risk-aversion measure (based on Gneezy and Potters, 1997), and an incentivized dictator game in which they could share the equivalent of 1 GBP in points with a randomly chosen participant of the experiment. Participants were also asked to briefly explain their strategy used in the network game. After finishing the survey, they received information about their final earnings, exited the study and received their payments online. We provide further information on the experimental sample and the survey questions in the Appendix.

We recruited 15 groups for each treatment with $N = 5$ and 10 groups for each treatment with $N = 9$.[13] This means that there were 420 participants in the study in total. Participants received 3.5 GBP as fixed show-up fee, in addition to their variable earnings in the 30 periods of network game and some of the survey questions that were incentivized. The payment from the network game was the sum of payoffs earned over the 30 periods[14], converting points to British pounds at the rate 150 points = 1 GBP. The average total payment was 12.11 GBP, including the 3.5 GBP show-up fee. The experiment was registered in the AEA RCT Registry (Horvath, 2022).

---

[13] We also had a number of groups who did not finish the experiment because one of the group members did not submit their decision within the 1-minute time limit given to the participants and hence, dropped out of the online experiment. This sort of dropout may happen due to inattention or issues with internet connections. Participants who dropped out received no payment, while the other participants in these incomplete groups received the show-up fee and their payoffs from the network game up to the period when the dropout happened. These rules were communicated clearly in the instructions before the start of the experiment. The number of incomplete groups by treatment were as follows: *N5_LowCost*: 2, *N5_HighCost*: 2, *N9_LowCost1*: 4, *N9_LowCost2*: 5, *N9_HighCost*: 5. The dropout rate of our experiment is similar to other online experiments involving groups (see Arechar et al., 2018). We compare behavior in the dropout groups to those that finished the experiment in the Appendix and find no significant difference.

## 4. Results on main outcomes

### 4.1 Network structure

We start the presentation of results by describing the network structures formed in the experiment. For each treatment, we count the relative frequency of each equilibrium network being played over the 30 periods of the experiment. We also compute the same statistic considering the last 10 periods only, in order to study the networks formed towards the end of the experiment when participants had more experience with the decisions. In addition, we report the relative frequency of network structures that differ in at most 2 links from the equilibrium network. This is to allow for the possibility that participants form the equilibrium network but sometimes make mistakes at the linking decision. Table 3 reports the findings.

The following results stand out. Considering the treatments with larger group size ($N = 9$), we find that the equilibrium networks are never formed. When we allow for a deviation of two links from the equilibrium networks, in a few cases the complete network is formed, but it is less than 10% of the cases in all treatments with $N = 9$. Regarding the treatments with smaller group size ($N = 5$), the complete network is the most focal outcome when there are multiple equilibria as in *N5_HighCost*, but it is only played with 18.7% frequency even in the last 10 periods. Its frequency goes up to 66.7% when we consider networks that deviate in at most 2 links from the equilibrium network. Regarding N5_LowCost, where the only equilibrium is the complete network, it is played only 25.3% of the times in the last 10 periods or 76.7% of the times when mistakes are allowed. Thus, even in this simplest case, the equilibrium network is not played with 100% frequency.

---

[14] In the Appendix, we study the impact of accumulated payoffs on behavior to understand whether individuals with higher payoffs accumulated over the first 20 periods stop linking to others because they are satisfied with their earnings. We do not find evidence of such satisficing behavior.

An interesting question is what network structures are formed instead of the equilibrium networks. Table 4 provides summary statistics of the network structures formed in the last 10 periods and compares them to the predictions under the most focal complete network.[15] Regarding the treatments with smaller group size ($N = 5$), we find that about 80% of all potential links are present in the network. The average degree is 3.343 in *N5_LowCost* and 3.323 in *N5_HighCost*, while the average smallest degree within a group is 2.74 in *N5_LowCost* and 2.52 in *N5_HighCost*. All these statistics are significantly lower than the prediction of 4 under the complete network, indicating that the network is less connected than predicted. In addition, we do not find any significant treatment differences in the network statistics between *N5_LowCost* and *N5_HighCost*, which indicates that lowering the linking cost from 3.9 to 1, does not make the network more connected.

Turning to the treatments with larger group size ($N = 9$), we find that 69.2% of all potential links are present in the network in *N9_LowCost1* and the same statistic is 60.1% and 78.8% for *N9_HighCost* and *N9_LowCost2*, respectively. These fractions are significantly lower than the predicted 1 under the complete network. The low connectivity of the network is also shown by the low average degree. The average degree is 5.544 in *N9_LowCost1*, 6.302 in *N9_LowCost2*, and 4.820 in *N9_HighCost*, all significantly lower than the predicted 8. Comparing the average degree between *N9_LowCost1* and *N9_HighCost*, we again find that lowering the linking costs has no significant effect on the connectivity of the network at the 5% significance level. In addition, we find substantial variation within group in terms of connectedness, the average minimum degree varies between 2.830 and 4.350 across the treatments, while the average maximum degree is between 6.960 and 7.830. There are about 4 links of difference between the least and most connected group members.

---

[15] In the Appendix we provide further information about the networks formed by depicting the network structures in the last 5 periods for each treatment and group (see Figure A8-A12). In addition, we show the evolution of the number of links and average degree in the network over the 30 periods of the experiment (see Figure A4). In these last graphs, we can see that the average degree does not change much in the last 10 periods of the experiment.

Next, we study the impact of group size on connectivity by comparing the treatments *N5_LowCost* and *N9_LowCost2*. Considering the total number of links and the average degree, we find significantly higher absolute connectivity in the larger network (see the test results in Table 4), which is not surprising since individuals have more chances to create links in the larger group. However, considering the fraction of links present relative to the complete network, we do not find statistically significant differences. 83.5% of all links are present in the network in *N5_LowCost* and the same fraction is 78.8% for *N9_LowCost2*. Our results thus suggest that the relative connectivity of the network does not depend on the group size which also shows the robustness of our results.

In sum, regarding RQ1, we find that individuals rarely form the equilibrium network, especially for the larger group size. In all treatments, the connectivity of the network is significantly lower than in the most focal equilibrium of complete network, which answers RQ1a and RQ1b. Regarding RQ3, we find that these results hold for all levels of linking costs and group size considered. In addition, these factors have no significant effects on the fraction of links present in the network.

**Table 3: Relative frequency of equilibrium network(s) formed by treatment**

| All periods | | | | | | |
|---|---|---|---|---|---|---|
| Treatment | Complete network | Complete network ±2 links | Empty network | Empty network ±2 links | Star network ±2 links | Star network ±2 links |
| *N5_LowCost* | 0.229 | 0.751 | | | | |
| *N5_HighCost* | 0.151 | 0.631 | 0 | 0 | 0 | 0.084 |
| *N9_LowCost1* | 0 | 0.003 | | | | |
| *N9_LowCost2* | 0 | 0.09 | | | | |
| *N9_HighCost* | 0 | 0 | 0 | 0 | 0 | 0 |
| Last 10 periods | | | | | | |
| Treatment | Complete network | Complete network ±2 links | Empty network | Empty network ±2 links | Star network ±2 links | Star network ±2 links |
| *N5_LowCost* | 0.253 | 0.767 | | | | |
| *N5_HighCost* | 0.187 | 0.667 | 0 | 0 | 0 | 0.107 |
| *N9_LowCost1* | 0 | 0 | | | | |
| *N9_LowCost2* | 0 | 0.08 | | | | |
| *N9_HighCost* | 0 | 0 | 0 | 0 | 0 | 0 |

*Note:* To compute the relative frequency, we count the number of times a given equilibrium network was formed and we divide it by the number of periods. We also consider the networks that differ in no more than 2 links from the equilibrium network, and count their relative frequency.

**Table 4: Network statistics in the last 10 periods by treatment and treatment comparisons**

| Treatment | Average number of links (std. dev) | Fraction of links relative to complete network (std. dev) | Average degree (std. dev) | Average minimum degree (std. dev) | Average maximum degree (std. dev) | Average clustering coefficients (std. dev) |
|---|---|---|---|---|---|---|
| *Prediction under complete network for N=5* | 10 | 1 | 4 | 4 | 4 | 1 |
| N5_LowCost | 8.353*** (1.262) | 0.835*** (0.126) | 3.343*** (0.502) | 2.740*** (0.782) | 3.887** (0.223) | 0.818*** (0.186) |
| N5_HighCost | 8.040*** (1.143) | 0.804*** (0.114) | 3.223*** (0.446) | 2.520*** (0.718) | 3.873** (0.219) | 0.800*** (0.142) |
| *Comparison of N5_LowCost vs. N5_HighCost, MW test, z-score (p-value)* | 0.954 (0.340) | 0.933 (0.351) | 0.933 (0.351) | 0.933 (0.351) | 0.083 (0.933) | 0.892 (0.373) |
| *Prediction under complete network for N=9* | 36 | 1 | 8 | 8 | 8 | 1 |
| N9_LowCost2 | 28.306*** (3.487) | 0.788*** (0.097) | 6.302*** (0.775) | 4.350*** (1.237) | 7.830* (0.295) | 0.827*** (0.080) |
| N9_LowCost1 | 24.927*** (4.047) | 0.692*** (0.112) | 5.544*** (0.890) | 3.600*** (0.967) | 7.645** (0.585) | 0.745*** (0.103) |
| N9_HighCost | 21.650*** (4.780) | 0.601*** (0.133) | 4.820*** (1.047) | 2.830*** (1.013) | 6.960** (1.033) | 0.636*** (0.162) |
| *Comparison of N9_LowCost1 vs. N9_HighCost, MW test, z-score (p-value)* | 1.760* (0.078) | 1.760* (0.078) | 1.760* (0.078) | 1.760* (0.078) | 1.127 (0.259) | 1.549 (0.121) |
| *Comparison of N5_LowCost vs. N9_LowCost2, MW test, z-score (p-value)* | -4.160*** (<0.001) | 1.442 (0.149) | -4.160*** (<0.001) | -3.051*** (0.002) | -4.160*** (<0.001) | 0.721 (0.471) |

*Note:* We compute network measures for each group and each period, including the number of links, fraction of links present in the network, average degree, minimum and maximum degree and clustering coefficient. We take averages of these over the last 10 periods for each group to form independent observations and compare them by one-sample Wilcoxon signed-rank tests to the predictions under the complete network equilibrium. Stars attached to the average numbers indicate significance levels of these tests: *** 1%, ** 5%, *10%. We also compare treatments by Mann-Whitney tests as indicated in the first column.

### 4.2 Effort choice and payoffs

Next, we turn to the analysis of the effort choice and welfare in the networks formed in the different treatments. Table 5 shows the average effort, average per period payoffs and the relative efficiency compared to the payoffs in the complete network with NE effort level. We report these statistics considering the last 10 periods of the experiment and compare them to the predictions under the most focal and Pareto-dominant complete network NE by non-parametric tests.[16] We also compare the effort levels realized in the experiment to the NE effort level *on the network that was formed in the experiment*, which we compute using the formula described in Result 1 in section 2.

**Table 5: Effort choice, payoffs and relative efficiency in the last 10 periods by treatment**

| Treatment | (1) Effort Prediction complete network NE | (2) Data Avg. effort (std. dev.) Compared to predictions | (3) Optimal average effort given the network structure in the experiment Compared to actual effort in the experiment | (4) Per period payoff Prediction complete network NE | (5) Data Avg. payoff (std. dev.) Compared to predictions | (6) Relative efficiency compared to equilibrium in complete network (realized payoff/equilibrium payoff) Compared to 1 |
|---|---|---|---|---|---|---|
| N5_LowCost | 4.167 | 4.470 (0.682) | 3.798** | 32.72 | 21.526*** (6.195) | 0.658*** (0.189) |
| N5_HighCost | 4.167 | 4.603 (1.217) | 3.729*** | 26.92 | 17.380*** (7.798) | 0.646*** (0.289) |
| N9_LowCost1 | 5 | 5.059 (1.029) | 3.894*** | 46 | 22.005*** (15.597) | 0.478*** (0.339) |
| N9_LowCost2 | 12.5 | 7.839*** (2.537) | 7.400 | 308.50 | 104.599*** (61.128) | 0.339*** (0.198) |
| N9_HighCost | 5 | 4.751 (0.678) | 3.650*** | 40 | 13.423*** (10.907) | 0.336*** (0.273) |

Note: We compute average effort, per period payoffs and relative efficiency compared to the complete network equilibrium for each group and each period. We take averages of these over the last 10 periods for each group to get independent observations and present the standard deviation of these group-level quantities in the parentheses. In column (2), (5) and (6), we compare the group-level averages to the equilibrium predictions under the complete network by one-sample Wilcoxon signed-rank tests. Stars attached to the average numbers indicate significance levels of these tests: *** 1%, ** 5%, *10%. In column (3), we report the NE average effort in the network formed in the experiment using the formula shown in Result 1 in section 2. Stars indicate the significance levels of one-sample Wilcoxon signed-rank tests comparing this NE prediction to the experimental data presented in column (2).

Regarding the choice of effort, we find that the average effort level is very close and not significantly different from the NE predictions under the complete network in four of the five treatments. The only exception is *N9_LowCost2* where the average effort is significantly lower than the NE predictions. The finding that the effort level is not significantly different from the predictions under the complete network NE in most of the treatments is surprising since the actual network structure is less connected than the complete network. This suggests that individuals overprovide effort relative to the *optimal value on the actual network* since in the less connected network that is formed in the experiment, the optimal effort should be lower as well. This conjecture is confirmed in column 3 of Table 5 which shows the NE effort levels *given the actual network structure formed in the experiment*. We can see the average effort chosen in the experiment is significantly higher than these NE effort levels in four out of the five treatments.[17] This result suggests that the overprovision of effort obtained in Gallo and Yan (2021) for fixed networks also emerges when the network connections are endogenously chosen by the individuals. We also note that while the effort provision is higher than the NE effort levels, it is significantly lower than the welfare-maximizing effort levels stated in Table 2.[18]

---

[16] In the Appendix, we show the evolution of average effort and relative efficiency compared to the equilibrium complete network over the 30 periods of the experiment (see Figure A5). We can see that these quantities change very little in the last 10 periods of the experiment.

Despite the overprovision of effort, we find substantial welfare losses in our experiment. We obtain in all five treatments that the average per period payoffs are significantly lower than predicted under the most focal and Pareto-dominant equilibrium of complete networks. We also compute the percentage of payoffs realized compared to this equilibrium, which is significantly lower than 1 in all treatments. Since payoffs do not reach even the equilibrium payoffs, it is certain that they are lower than the payoffs in the efficient allocation given that the equilibrium is not efficient. The main reason for the loss of welfare is the lower connectivity of the network compared to the equilibrium, which limits the welfare-improving positive spillover effects among the effort choices of individuals. However, there are some other effects connected to the linking decisions as well, which we discuss in more detail in the Appendix.

Turning to the treatment comparisons, we find by comparing *N5_LowCost to N9_LowCost2* that the welfare loss relative to the equilibrium payoffs increases with the group size. In *N5_LowCost*, 65.8% of the equilibrium payoffs are realized, the same number is only 33.9% for *N9_LowCost2.* The difference is strongly statistically significant (MW-test, z-score: 3.106, p-value: 0.002). While in both networks about 20% of the links are missing, in absolute terms, more links are missing from the network of the larger group, which limits the welfare-improving effects of positive spillovers. We thus find that the deviation from the equilibrium is more costly in terms of welfare reduction in the larger group.[19]

---

[17] We note that it is not general to find overprovision of effort in games with strategic complementarity. For example, Antinyan et al. (2020) and Charness et al. (2014) do not find deviation from equilibrium choices in network games. Similarly, the experimental results on Bertrand competition with differentiated products shows that collusion on higher prices happens only in specific settings (Engel, 2007; Potters and Suetens, 2013).
[18] Statistical tests all yield p-values smaller than 0.001 and are thus omitted for brevity.

In sum, regarding RQ2, we find that in four of the five treatments the average effort is significantly higher than the NE effort level on the realized network structure. However, the overprovision of effort does not lead to welfare improvements, as we obtain a significant welfare loss compared to the equilibrium and efficient allocations in all treatments, stemming from the low connectivity of the network formed. As for RQ3, we find that the extent of the welfare loss is bigger in the larger groups.

## 5. Decision-making at the individual level

### 5.1 Effort choice and overprovision of effort

In this section, we aim to explain the aggregate outcomes presented in section 4 by analyzing the individual behavior. We start by modeling the individual effort choices in order to understand the sources of overprovision of effort and why it happens in only four and not all treatments. Figure A5 shows the evolution of effort over the 30 periods. We can notice that in the four treatments with overprovision of effort, the average effort starts above the equilibrium value and slowly decreases, while in the fifth treatment, it starts below it and slowly increases. The dynamics towards the equilibrium suggests that the myopic best-response dynamics is a fitting model for the effort choice. According to this model, individuals give best response to the belief that their opponents will repeat the same strategy as in the previous period. This leads to the following equation for the effort choice that is derived from the payoff function:

$$x_{it}^*(x_{k,t-1}) = \frac{\theta + \lambda \sum_{k \in N_i(G_{t-1})} x_{k,t-1}}{\beta}$$

---

[19] This finding holds even if the average effort (MW-test, z-score: -3.439, p-value: 0.001) and absolute payoffs (MW-test, z-score: -4.160, p-value: 0.001) are significantly larger in N9_LowCost2 than in N5_LowCost, which is due to the larger absolute number of nodes and links in the larger group.

where the term $\sum_{k \in N_i(G_{t-1})} x_{k,t-1}$ sums over the previous effort choices of network neighbors. However, this model converges to the NE over time, while our data shows deviations from it. To capture the overprovision of effort, we add two terms to the model: 1) inertia meaning that individuals are influenced by their own choice in the previous period, 2) conformity or imitation meaning that individuals are influenced by others in the group who are not their neighbors (and are thus not part of the myopic best-response calculation). These two factors slow down the convergence to the equilibrium.

Based on these considerations, we propose the following model of effort choice:

$$x_{it} = \beta_0 x_{i,t-1} + \beta_1 x_{it}^* + \beta_2 \sum_{k \in N_{-i}/N_i(G_{t-1})} x_{k,t-1} + \varepsilon_{it}$$

where the second term is the myopic best-response effort choice, the third term is the sum of effort choices of non-neighbors in the previous period and the last is an error term.

We estimate this model separately for each treatment and report the results in Table 6.[20] We can see that the coefficient of the best-response effort level is positive and statistically significant for all treatments, as is the coefficient of the lagged dependent variable capturing inertia. This shows that myopic best-response with inertia is a valid model to describe the effort choice. Interestingly, the coefficient of the sum of non-neighbors' effort choices is positive and statistically significant only in the treatments with overprovision of effort but not in *N9_LowCost2* where the effort level converges to the NE starting from a lower value. When effort in the group is above the NE value, individuals positively reciprocate the effort of both the neighbors (via the best-response term) and the non-neighbors. This is a beneficial behavior since payoffs are higher if all group members jointly increase their effort above the NE due to the positive externalities of effort. In contrast, when effort levels are below the NE as in *N9_LowCost2*, there is no such payoff advantage, individuals thus do not reciprocate the effort levels of non-neighbors, they only give best response to the choices of neighbors which increases the speed of convergence. These effects explain the difference in behavior between *N9_LowCost2* and the other four treatments.

---

[20] Figure A6 plots the fitted dependent variable of this model against the experimental data, and shows that the model produces considerably good fit of the data.

**Table 6: Estimation of myopic-best response model of effort choice**

| Dependent variable: Effort | (1) | (2) | (3) | (4) | (5) |
|---|---|---|---|---|---|
| Sample | N5_LowCost | N5_HighCost | N9_LowCost1 | N9_HighCost | N9_LowCost2 |
| Own effort in t-1 | 0.090** | 0.161*** | 0.089** | 0.298*** | 0.324*** |
|  | (0.036) | (0.058) | (0.039) | (0.072) | (0.056) |
| Best-response effort to neighbors' choice | 0.966*** | 0.900*** | 0.455*** | 0.763*** | 0.376*** |
|  | (0.040) | (0.069) | (0.066) | (0.067) | (0.046) |
| Sum of non-neighbors effort in t-1 | 0.085*** | 0.036*** | 0.019*** | 0.018*** | 0.014 |
|  | (0.018) | (0.012) | (0.007) | (0.004) | (0.009) |
| Observations | 2,248 | 2,250 | 2,700 | 2,700 | 2,700 |

*Note:* Multi-level mixed-effects regressions with standard errors clustered at the group level (reported in parenthesis). Dependent variable: individual effort choice. Stars indicate significance levels: *** 1%, ** 5%, *10%.

## 5.2 Linking decision and the reasons for the lack of links

In this section, we study the linking decisions at the individual level with the aim to understand why individuals do not form the Pareto-dominant complete network that would result in the highest payoffs. We analyze the following hypotheses in detail: 1) the missing links are not beneficial to form; 2) the missing links are beneficial but bring negligible benefits; 3) individuals do not understand the benefits from the links and hence do not create them; 4) individual make linking decisions based on relative position: they form links to top effort providers in the group and dismiss links to relative low effort providers. For the sake of brevity, we relegate the testing of further hypotheses to the Appendix, these include: 1) individuals strategically do not link to others in order to save the costs of initiating links; 2) individuals punish others by withdrawing links to them; 3) missing links towards the end of the experiment can be explained by satisficing behavior of those who earned high payoffs early on in the experiment.

Theoretically, when deciding about linking, individuals need to compare the costs and benefits of establishing a link. The benefits of establishing a link between individual $i$ and $j$ are $\lambda x_i x_j$, which depend on the effort choices of both individuals, this is compared to the linking cost $\kappa$. In the Appendix, we show that the vast majority of missing links could indeed be profitably added as the benefits would exceed the linking costs (see Table A5). For example, in *N5_LowCost* 98.7% of the missing links would bring positive benefits, the same number is 72.6% in *N5_HighCost*. Regarding the size of the benefits of missing links, we depict on Figure A8 the average benefit that a missing link would bring and compare it to the average per period payoffs of the participants. We can see that on average, an added missing link would increase the per period payoffs of an individual by about 20-50%, which we believe cannot be labeled as negligible. We thus have shown evidence against our first two hypotheses stating that the missing links are not beneficial or the benefits of link formation are negligible.

While adding links would be beneficial, it may be the case that individuals do not understand the benefits (hypothesis 3). To dig deeper in this explanation, we first study the relationship between link formation and the benefits of linking. As mentioned above, the benefits of a link depend on the effort choices of the two individuals involved in the link. We compute these benefits based on the effort choices in the previous period, so that the benefits of a link between individual $i$ and individual $j$ in period $t$ become $\lambda x_{i,t-1} x_{j,t-1} - \kappa$.[21] We study the impact of link benefits on initiating a link by running logistic regressions at the link level where the binary dependent variable takes the value of 1 if an individual initiated a link to a given another group member and the value of 0 if she did not. We apply individual random-effects logistic regressions and cluster standard errors at the individual level. We pool the data from all periods and treatments and control for treatment differences by adding the parameter values by which the treatments differ (group size, cost of linking and the $\lambda$ parameter of the payoff function) to the regressions. We report odds ratios in Table 7.

In column (1) of Table 7, we include the benefits of the link as a dummy variable, which takes the value of 1 when the benefit is positive and the value of 0 otherwise. We can see that beneficial links are more likely to be formed than the non-beneficial ones, as the odds ratio of this dummy variable is statistically significant and above 1. In column (2), we add the size of the link benefit and find that individuals are more like to form those links that bring higher benefits. In column (3), we break down the benefits into four components: the individual's own effort level, the partner's effort level to whom the link goes, the linking cost and the $\lambda$ parameter that multiplies the two effort levels. We can see that out of the four components, only the partner's effort level is statistically significant at the 5% level, with the expected sign. We thus find that individuals take into account the benefits of a link when making linking decisions, however, they approximate the value of a link by the partner's effort choice in the previous period.

---

[21] We use the effort choices in the previous period because these are observable to the experimental participants when making decisions and this is consistent with the myopic best-response model introduced in section 5.1.

**Table 7: The impact of link benefits on link formation**

| Dependent variable: Link initiated (1=Yes, 0=No) | (1) | (2) | (3) |
|---|---|---|---|
| Link initiated in t-1 (1=Yes, 0=No) | 2.851*** | 2.827*** | 2.800*** |
|  | (0.188) | (0.186) | (0.182) |
| Link was beneficial in t-1 (1=Yes, 0=No) | 1.284*** |  |  |
|  | (0.068) |  |  |
| Amount of link benefit in t-1 |  | 1.015*** |  |
|  |  | (0.003) |  |
| Own effort in t-1 |  |  | 1.004 |
|  |  |  | (0.006) |
| Partner's effort in t-1 |  |  | 1.083*** |
|  |  |  | (0.009) |
| Lambda |  |  | 5.827 |
|  |  |  | (6.252) |
| Linking cost |  |  | 0.909* |
|  |  |  | (0.045) |
| Large group (1=Yes, 0=No) | 0.620*** | 0.595*** | 0.623*** |
|  | (0.067) | (0.063) | (0.113) |
| Constant | 0.649*** | 0.729*** | 0.342** |
|  | (0.066) | (0.067) | (0.164) |
| Observations | 84,948 | 84,948 | 84,948 |

*Note:* Individual random-effects regressions at the link level with standard errors clustered at the individual level (reported in parenthesis). Odds ratios are reported. Dependent variable: initiating a link (1 if the individual initiated a link to a given opponent, 0 otherwise). Stars indicate significance levels: *** 1%, ** 5%, *10%. Sample: all treatments.

By focusing on the opponent's effort level when deciding about linking, individuals may misjudge the linking benefits and make suboptimal decisions. To explore if this explains the lack of links, we run an additional experimental treatment where we make the benefits of each link explicit to the individuals. We build on treatment *N9_LowCost1* and modify the feedback page shown after each period by adding the information on "the additional number of Points that could be earned by establishing a link to each of the other group members". We calculate the value of a link to each group member and display it on the feedback page in the table that lists every other group member's effort level and network neighbors. An example of the feedback page is shown on Figure A3 in the Appendix. Individuals will thus have accurate information on the benefits of links in this treatment, and we expect more links to be formed than in *N9_LowCost1*, which does not have such information.

We obtain the following results. The average degree in the last 10 periods was 5.544 in *N9_LowCost1* and it is 5.597 in the additional treatment[22], the two averages are not statistically significantly different (MW-test, z-score: -0.082, p-value: 0.953). In addition, we obtain that the average effort levels in the last 10 periods are also not statistically significantly different between the two treatments: the average effort level was 5.059 in *N9_LowCost1* and it is 5.480 in the additional treatment (MW-test, z-score: -0.163, p-value: 0.905). We thus do not find any impact of showing information on link benefits to the individuals on the key outcomes of the experiment.

Finally, we test our fourth hypothesis according to which individuals use the relative effort ranking of the partner within the group to decide to whom to extend a link. The idea is that individuals may link to the top effort providers in the group who bring the highest benefits and ignore other beneficial links that would go to others with lower relative effort ranking. To analyze this, we run logistic regressions of link formation and include the relative ranking of the partner in terms of effort provision in the group in period $t-1$ as an explanatory variable. In treatments with $N=5$, we take rank 3 (the median) effort as the reference category, while in treatments with $N=9$, the reference category is being ranked as 4th, 5th or 6th in the group. Then, we create dummy variables capturing whether the partner's effort in $t-1$ is below or above the reference category. In addition, to check the robustness of the results to the reference category, we run regressions using a dummy variable measuring whether the partner's effort is above or below the group's average effort level. We add these variables to the regressions besides the absolute value of the partner's effort[23] in $t-1$, the lagged dependent variable and the treatment dummy variables. The regression results are reported in Table 8.

---

[22] The standard deviations across groups are 0.892 in *N9_LowCost1* and 1.137 in the additional treatment.
[23] We also include the opponent's effort level in absolute value in the regression, to make sure that the relative effort variables do not merely capture the opponent's effort provision.

In all regressions, we can observe that the partner's relative position in the group matters on top of the absolute value of effort provided by the partner. Individuals are more likely to link to a partner if her effort is above the group median effort level and less likely if it is below the group median (see columns 1-3). Similarly, individuals are less likely to link to a partner if her effort level is below the group average effort level compared to being above the group average (see columns 4-6). Individuals thus use the relative position of the partner in the group in terms of effort provision to judge the benefit of the link and do not link to relative low effort providers even if that link would bring positive absolute benefits.[24]

### Table 8: Relative position in the group and linking

| Dependent variable: Link initiated=1 (no=0) | (1) | (2) | (3) | (4) | (5) | (6) |
|---|---|---|---|---|---|---|
| Reference category | Median | Median | Median | Average | Average | Average |
| Sample | All treatments | Treatments with N=5 | Treatments with N=9 | All treatments | Treatments with N=5 | Treatmens with N=9 |
| Link initiated in t-1 (1=yes, 0=no) | 2.794*** (0.182) | 2.176*** (0.223) | 2.993*** (0.233) | 2.789*** (0.181) | 2.173*** (0.222) | 2.988*** (0.233) |
| Opponent's effort in t-1 | 1.047*** (0.009) | 1.035** (0.014) | 1.046*** (0.010) | 1.051*** (0.008) | 1.034*** (0.013) | 1.053*** (0.009) |
| Above group median effort in t-1 | 1.281*** (0.046) | 1.158** (0.076) | 1.326*** (0.058) | | | |
| Below group median effort in t-1 | 0.926** (0.029) | 0.903* (0.051) | 0.921** (0.034) | | | |
| Below average effort in t-1 | | | | 0.770*** (0.027) | 0.781*** (0.046) | 0.764*** (0.033) |
| N5_HighCost | 0.828 (0.140) | 0.820 (0.143) | | 0.821 (0.138) | 0.813 (0.142) | |
| N9_LowCost1 | 0.547*** (0.086) | | | 0.538*** (0.085) | | |
| N9_HighCost | 0.412*** (0.068) | | 0.755* (0.109) | 0.406*** (0.067) | | 0.756* (0.109) |
| N9_LowCost2 | 0.725* (0.136) | | 1.326* (0.226) | 0.707* (0.133) | | 1.305 (0.222) |
| Constant | 0.687*** (0.093) | 0.873 (0.136) | 0.363*** (0.043) | 0.815 (0.110) | 1.007 (0.154) | 0.423*** (0.050) |
| Observations | 84,948 | 17,996 | 66,952 | 84,948 | 17,996 | 66,952 |

*Note:* Individual random-effects regressions at the link level with standard errors clustered at the individual level (reported in parenthesis). Odds ratios are reported. Dependent variable: initiating a link (1 if the individual initiated a link to a given opponent, 0 otherwise). Stars indicate significance levels: *** 1%, ** 5%, *10%. Sample: all treatments in columns 1 and 4, treatments with $N = 5$ in columns 2 and 5, treatments with $N = 9$ in columns 3 and 6.

---

[24] This phenomenon may also explain why linking costs have no significant impact on the number of links formed. Linking to top effort providers is the most beneficial thing to do independently of the level of linking costs.

Interestingly, we find the same phenomenon in the additional treatment where we show the correct value of link benefits to the participants of the experiment. In the Appendix (see Table A6), we demonstrate that individuals tend to link to top effort providers in the group and do not establish links to relative low effort providers even if they are aware that those links would increase their payoffs.

Our analysis in this section thus finds that creating the missing links would result in considerable benefits for the individuals. Participants are more likely to form the links that bring higher benefits. Misunderstandings of those benefits cannot explain the missing links since the number of links does not increase if we explicitly show the linking benefits to the individuals. Instead, we find that individuals base their linking decisions on the relative ranking of the partner within the group in terms of effort provision. This leads to missing links as individuals do not form the links to relative low effort providers, even if those links would be beneficial in absolute terms. Additional analysis in the Appendix shows that strategic considerations, satisficing behavior and punishment of those decreasing effort provision cannot provide additional explanations of missing links.

## 6. Conclusions

In this paper, we study network formation and effort provision in a game of positive externality and complementarity. Individuals initiate links at a cost and choose a costly effort that benefits themselves and their direct neighbors in the network. The payoffs are determined by the linear-quadratic utility function. Complementarities imply that individuals have incentives to set a higher effort level if their neighbors provide higher effort as well. Positive externalities imply that the Nash equilibrium effort level and network structure are typically not efficient. In particular, in the efficient allocation individuals provide higher effort than the equilibrium level.

In this setting, we study whether individuals will form a Nash equilibrium network and choose the Nash equilibrium effort level or will be able to coordinate on a higher effort level that increases the group's welfare. We also study equilibrium selection in our treatments with three equilibria. In addition, we vary the group size and the cost of linking, and analyze their impact on the network formed and the efficiency of the outcomes.

We find that experimental participants rarely form the equilibrium network. In all treatments, the networks formed in the experiment are under-connected on average relative to the equilibrium complete network that is payoff-dominant. In terms of effort provision, in four of our five treatments, individuals choose a higher effort level on average compared to the equilibrium effort on the network that they form. We use a modified version of the myopic best-response model to explain individual effort provision decisions. However, the higher effort provision does not lead to higher payoffs due to the lack of links in the networks formed. The lack of links also limits positive spillovers of effort provision between connected individuals which would further increase effort provision and welfare. The welfare realized in the experiment is lower than the welfare in the Nash equilibrium and, consequently, the efficient allocation. The welfare loss is higher in magnitude in the larger groups. We find that individuals tend to connect to the top effort providers in the group and ignore otherwise beneficial links to relative low effort providers in the group. This explains the lack of links in the network.

Our results have several implications. Firstly, we obtain that individuals do not exploit all potential benefits of collaboration, because they do not link to everyone whom they could profitably link to. We find that only about 60-80% of all potential links are present in the networks. This result calls for policy interventions that increase the connectedness of the network in systems of networked collaborations. Our results suggest that decreasing the linking costs may not be sufficient to achieve this goal. Alternative instruments, such as increasing the benefits of link formation or increasing group cohesion among the individuals may result in better outcomes with higher welfare. Secondly, network formation in the linear-quadratic game has the potential to explain why we observe core-periphery networks in many applications (see e.g., Fricke and Lux, 2015; Juhász et al., 2020; König et al. 2012; Leydersdorff and Wagner, 2008). However, we find that the star network is never played in our treatments where it is an equilibrium. This suggests that strategic interactions among ex-ante homogenous players alone may not be able to explain the emergence of core-periphery networks, despite this being suggested by the theoretical models (Hiller, 2017; König et al. 2014). Heterogeneity among the agents may need to be added to the model to explain the emergence of core-periphery networks (a conclusion supported by the model in Int'l Veld et al., 2020). We leave these areas for future research.

## References

1. Amir, O., & Rand, D. G. (2012). Economic games on the internet: The effect of $1 stakes. PloS one, 7(2), e31461.
2. Antinyan, A., Horvath, G., & Jia, M. (2020). Positional concerns and social network structure: An experiment. *European Economic Review*, *129*, 103547.
3. Arifovic, J., & Ledyard, J. (2018). Learning to alternate. *Experimental Economics*, *21*, 692-721.
4. Baetz, O. (2015). Social activity and network formation. *Theoretical Economics*, *10*(2), 315-340.
5. Bala, V., & Goyal, S. (2000). A noncooperative model of network formation. *Econometrica*, *68*(5), 1181-1229.
6. Ballester, C., Zenou, Y., & Calvó-Armengol, A. (2010). Delinquent networks. *Journal of the European Economic Association*, *8*(1), 34-61.
7. Berninghaus, S. K., Ehrhart, K. M., & Keser, C. (2002). Conventions and local interaction structures: experimental evidence. *Games and Economic Behavior*, 39(2), 177-205.
8. Berninghaus, S. K., Ehrhart, K. M., Ott, M., & Vogt, B. (2007). Evolution of networks—an experimental analysis. *Journal of Evolutionary Economics*, *17*(3), 317-347.
9. Bramoullé, Y., Kranton, R., & D'amours, M. (2014). Strategic interaction and networks. *American Economic Review*, *104*(3), 898-930.
10. Brañas-Garza, P., García-Muñoz, T., & González, R. H. (2012). Cognitive effort in the beauty contest game. *Journal of Economic Behavior & Organization*, *83*(2), 254-260.
11. Brañas-Garza, P., Kujal, P., & Lenkei, B. (2019). Cognitive reflection test: Whom, how, when. *Journal of Behavioral and Experimental Economics*, *82*, 101455.
12. Callander, S., & Plott, C. R. (2005). Principles of network development and evolution: An experimental study. *Journal of Public Economics*, *89*(8), 1469-1495.
13. Calvó-Armengol, A., Patacchini, E., & Zenou, Y. (2009). Peer effects and social networks in education. *The review of economic studies*, *76*(4), 1239-1267.
14. Caria, A. S., & Fafchamps, M. (2019). Expectations, network centrality, and public good contributions: Experimental evidence from India. *Journal of Economic Behavior & Organization,* 167, 391-408.
15. Caria, A. S., & Fafchamps, M. (2020). Can people form links to efficiently access information?. *The Economic Journal*, *130*(631), 1966-1994.
16. Carpenter, J., Graham, M., & Wolf, J. (2013). Cognitive ability and strategic sophistication. *Games and Economic Behavior*, *80*, 115-130.
17. Cassar, A. (2007). Coordination and cooperation in local, random and small world networks: Experimental evidence. Games and Economic Behavior, 58(2), 209-230.
18. Charness, G., Feri, F., Meléndez-Jiménez, M. A., & Sutter, M. (2014). Experimental games on networks: Underpinnings of behavior and equilibrium selection. *Econometrica*, *82*(5), 1615-1670.
19. Chen, D. L., Schonger, M., & Wickens, C. (2016). oTree—An open-source platform for laboratory, online, and field experiments. *Journal of Behavioral and Experimental Finance*, 9, 88–97.
20. Choi, S., Goyal, S., & Moisan, F. (2022). Connectors and influencers. Working Paper.
21. Choi, S., Kariv, S., & Gallo, E. (2016). Networks in the Laboratory. In *The Oxford Handbook of the Economics of Networks*.
22. Dawid, H., & Hellmann, T. (2014). The evolution of R&D networks. *Journal of Economic Behavior & Organization*, *105*, 158-172.
23. Ductor, L., Fafchamps, M., Goyal, S., & Van der Leij, M. J. (2014). Social networks and research output. *Review of Economics and Statistics*, *96*(5), 936-948.

24. Ductor, L., Goyal, S., & Prummer, A. (2021). Gender and collaboration. *The Review of Economics and Statistics*, 1-40.
25. Engel, C. (2007). How much collusion? A meta-analysis of oligopoly experiments. *Journal of Competition Law and Economics*, 3(4), 491-549.
26. Essers, D., Grigoli, F., & Pugacheva, E. (2022). Network effects and research collaborations: Evidence from IMF Working Paper co-authorship. *Scientometrics*, 1-24.
27. Falk, A., & Kosfeld, M. (2012). It's all about connections: Evidence on network formation. *Review of Network Economics*, *11*(3).
28. Fatas, E., Meléndez-Jiménez, M. A., & Solaz, H. (2010). An experimental analysis of team production in networks. *Experimental Economics*, 13(4), 399-411.
29. Frederick, S. (2005). Cognitive reflection and decision making. *Journal of Economic Perspectives*, 19(4), 25-42.
30. Fricke, D., & Lux, T. (2015). Core–periphery structure in the overnight money market: evidence from the e-mid trading platform. *Computational Economics*, *45*(3), 359-395.
31. Galeotti, A., & Goyal, S. (2010). The law of the few. *American Economic Review*, *100*(4), 1468-92.
32. Gallo, E., & Yan, C. (2015). The effects of reputational and social knowledge on cooperation. *Proceedings of the National Academy of Sciences*, *112*(12), 3647-3652.
33. Gallo, E., & Yan, C. (2021). Efficiency and equilibrium in network games: An experiment. *The Review of Economics and Statistics*, 1-44.
34. Gneezy, U., & Potters, J. (1997). An experiment on risk taking and evaluation periods. *The Quarterly Journal of Economics*, 112(2), 631-645.
35. Goeree, J. K., Riedl, A., & Ule, A. (2009). In search of stars: Network formation among heterogeneous agents. *Games and Economic Behavior*, *67*(2), 445-466.
36. Goyal, S., & Moraga-Gonzalez, J. L. (2001). R&d networks. *Rand Journal of Economics*, 686-707.
37. He, Simin, and Jiabin Wu. "Compromise and coordination: An experimental study." *Games and Economic Behavior* 119 (2020): 216-233.
38. Hiller, T. (2013) *Peer effects in endogenous networks.* Theoretical Economics (TE/2013/564). Suntory and Toyota International Centres for Economics and Related Disciplines, London, UK.
39. Horvath, G. (2022). Network formation and efficiency in linear-quadratic games: An experimental study. *AEA RCT Registry*. September 02. https://www.socialscienceregistry.org/trials/9999
40. Hsieh, C. S., Konig, M. D., Liu, X., & Zimmermann, C. (2018). Superstar Economists: Coauthorship networks and research output. *Available at SSRN 3266432*.
41. In't Veld, D., Van der Leij, M., & Hommes, C. (2020). The formation of a core-periphery structure in heterogeneous financial networks. *Journal of Economic Dynamics and Control*, 119, 103972.
42. Jadidi, M., Karimi, F., Lietz, H., & Wagner, C. (2018). Gender disparities in science? Dropout, productivity, collaborations and success of male and female computer scientists. *Advances in Complex Systems*, *21*(03n04), 1750011.
43. Juhász, S., Tóth, G., & Lengyel, B. (2020). Brokering the core and the periphery: Creative success and collaboration networks in the film industry. *Plos one*, *15*(2), e0229436.
44. Kiss, H. J., Rodriguez-Lara, I., & Rosa-García, A. (2016). Think twice before running! Bank runs and cognitive abilities. *Journal of Behavioral and Experimental Economics*, *64*, 12-19.
45. König, M. D., Battiston, S., Napoletano, M., & Schweitzer, F. (2012). The efficiency and stability of R&D networks. *Games and Economic Behavior*, *75*(2), 694-713.
46. König, M. D., Tessone, C. J., & Zenou, Y. (2014). Nestedness in networks: A theoretical model and some applications. *Theoretical Economics*, *9*(3), 695-752.
47. Kroher, M., & Wolbring, T. (2015). Social control, social learning, and cheating: Evidence from lab and online experiments on dishonesty. *Social Science Research,* 53, 311–324.
48. Lee, L. F., Liu, X., Patacchini, E., & Zenou, Y. (2021). Who is the key player? A network analysis of juvenile delinquency. *Journal of Business & Economic Statistics*, *39*(3), 849-857.
49. Leydesdorff, L., & Wagner, C. S. (2008). International collaboration in science and the formation of a core group. *Journal of informetrics*, *2*(4), 317-325.
50. Lindenlaub, I., & Prummer, A. (2021). Network structure and performance. *The Economic Journal*, *131*(634), 851-898.

51. Lindquist, M. J., & Zenou, Y. (2019). Crime and networks: Ten policy lessons. *Oxford Review of Economic Policy*, *35*(4), 746-771
52. Newman, M. E. (2004). Coauthorship networks and patterns of scientific collaboration. *Proceedings of the national academy of sciences*, *101*(suppl_1), 5200-5205.
53. Palan, S., & Schitter, C. (2018). Prolific. ac—A subject pool for online experiments. *Journal of Behavioral and Experimental Finance*, 17, 22–27.
54. Paolacci, G., Chandler, J., & Ipeirotis, P. G. (2010). Running experiments on amazon mechanical turk. *Judgment and Decision Making*, 5(5), 411–419.
55. Potters, J., & Suetens, S. (2013). Oligopoly experiments in the current millennium. A Collection of Surveys on Market Experiments, 51-74.
56. Rand, D. G., Nowak, M. A., Fowler, J. H., & Christakis, N. A. (2014). Static network structure can stabilize human cooperation. *Proceedings of the National Academy of Sciences*, 111(48), 17093-17098.
57. Rand, D. G. (2012). The promise of Mechanical Turk: How online labor markets can help theorists run behavioral experiments. *Journal of Theoretical Biology*, 299, 172–179.
58. Rand, D. G., Arbesman, S., & Christakis, N. A. (2011). Dynamic social networks promote cooperation in experiments with humans. *Proceedings of the National Academy of Sciences*, *108*(48), 19193-19198.
59. Riedl, A., Rohde, I. M., & Strobel, M. (2016). Efficient coordination in weakest-link games. *The Review of Economic Studies*, 83(2), 737-767.
60. Rong, R., & Houser, D. (2015). Growing stars: A laboratory analysis of network formation. *Journal of Economic Behavior & Organization*, *117*, 380-394.
61. Rosenkranz, S., & Weitzel, U. (2012). Network structure and strategic investments: An experimental analysis. Games and Economic Behavior, 75(2), 898-920.
62. Suri, S., & Watts, D. J. (2011). Cooperation and contagion in web-based, networked public goods experiments. PloS One, 6(3), e16836.
63. Vega-Redondo, F. (2016). Links and actions in interplay. In Bramoullé, Y., Galeotti, A., & Rogers, B. W. (Eds.). (2016). *The Oxford Handbook of the Economics of Networks*. Oxford University Press.

# Online Appendix to "Network formation and efficiency in linear-quadratic games: An experimental study"

Gergely Horvath[25]

*Division of Social Sciences, Duke Kunshan University*

## 1. Post-experimental survey and sample

Our post-experimental survey asks about basic demographic characteristics, including gender, age, highest education, nationality, student and work statuses. We also ask participants about the strategy they applied in the network formation game using an open-ended question. We elicit risk preferences using the incentivized investment task based on Gneezy and Potters (1997). Participants were presented with the following scenario: *"Imagine you have 100 points. You can invest some of it in a risky investment. If you invest Y in the risky investment, the investment return will be either 2.5*Y or 0 with equal probability. You may think of this as if the investment return was decided by a coin toss. The amount not invested in the risky investment is 100-Y. Please, enter below how much you would like to invest in the risky investment. After this, the computer will toss a coin and determine the investment return. Your payoff will be 100-Y+the investment return in points. This will be added to the payoffs you earned in Part 1 of the Experiment. The amount you would like to invest in the risky investment is:"* Note that risk-neutral individuals should invest all their endowment in the risky asset, while risk-averse individuals should invest less, depending on their degree of risk aversion.

In addition, we elicit the social preferences of the participants using a dictator game where they can share 150 Points, the equivalent of 1 GBP, with a randomly chosen another participant of the experiment who participated in another group. The text of the dictator game states: *"You receive an additional 150 Points. Your task is to decide how to allocate it between yourself and a randomly chosen participant of this experiment from another group (thus someone whom you have not interacted with). You can allocate these Points in any way you want. Remember, you are entitled to keep the Points for yourself. The amount you keep will be added to your earnings from the*

---

[25] Division of Social Sciences, Duke Kunshan University. Address: Division of Social Sciences, Duke Kunshan University, No. 8 Duke Avenue, Kunshan, Jiangsu Province, China 215316. E-mail address: horvathgergely@gmail.com. Tel: +86 155-01683821.

This research was supported by the National Natural Science Foundation of China (NSFC), under the 2021 NSFC Research Fund for International Excellent Young Scientists (RFIS-II) program, grant nr. 72150610501.

*experiment. The rest will be added to the earnings of a randomly chosen participant of this experiment from another group."*

We also ask the participants to fill out a cognitive reflection test consisting of three questions based on Frederick (2005). Each correctly answered question was rewarded with 30 Points, the equivalent of 0.2 GBP. The three questions read as follows:

- "A bat and a ball cost $1.10 in total. The bat costs $1.00 more than the ball. How much does the ball cost? (in cents)"
- "If it takes 5 machines 5 minutes to make 5 widgets, how long would it take 100 machines to make 100 widgets? (in minutes)"
- "In a lake, there is a patch of lily pads. Every day, the patch doubles in size. If it takes 48 days for the patch to cover the entire lake, how long would it take for the patch to cover half the lake? (in days)"

Earnings from all incentivized questions of the post-experimental survey were added to the participants' earnings from the network formation game.

Table A1 shows the main characteristics of the experimental sample. We can see that 48.6% of the sample is female, 48.8% of the sample is student, the average age is 27.631, somewhat higher than it is typical in lab experiments carried out with students at a university. Individuals completed about 16 years of education on average, which corresponds to college education.

**Table A1 shows main characteristics of the experimental sample.**

| Individual characteristics | Mean (std. dev.) |
|---|---|
| Gender (1=Female, 0=Male) | 0.486 (0.505) |
| Age | 27.631 (7.456) |
| Education in years | 16.075 (1.655) |
| Student status | 0.488 (0.500) |
| Risk aversion (amount invested in the risky asset) | 47.757 (28.709) |
| Amount kept in the dictator game out of 150 | 121.919 (32.421) |
| Number of correctly answered questions in the cognitive reflection test | 1.667 (1.253) |

References

1. Frederick, S. (2005). Cognitive reflection and decision making. *Journal of Economic Perspectives*, 19(4), 25-42.
2. Gneezy, U., & Potters, J. (1997). An experiment on risk taking and evaluation periods. *The Quarterly Journal of Economics*, 112(2), 631-645.

## 2. Robustness checks

### 2.1 Dropout groups

In this section, we analyze whether individual behavior is significantly different in the groups that dropped out during the experiment compared to those that finished the experiment. The number of dropout groups by treatment are as follows: The number of incomplete groups by treatment were as follows: *N5_LowCost*: 2, *N5_HighCost*: 2, *N9_LowCost1*: 4, *N9_LowCost2*: 5, *N9_HighCost*: 5. The periods when dropout happened are listed as follows: *N5_LowCost*: (2, 10), *N5_HighCost*: (6, 28), *N9_LowCost1*: (18, 9, 8, 26), *N9_LowCost2*: (11, 19, 17, 25, 24), *N9_HighCost*: (16, 13, 3, 27, 17).

We pool the data from complete and dropout groups and introduce a dummy variable which takes the value of 1 if the group dropped out, and the value of 0 otherwise. We use regression analysis to study whether this dummy variable capturing dropout has a significant correlation with decisions, including effort choice, number of neighbors initiated, number of neighbors formed and per period payoffs. Obviously, we have data from dropout groups only up to the period when the dropout happened. In addition, we cannot control for individual variables since dropout groups did not fill out the survey at the end of the experiment. Table A2 shows the results. We find that the coefficient of the dummy variable representing dropout groups is not statistically significant in any of the regressions. This means that individuals in dropout groups do not exhibit distinct behavior from those in groups that finished the experiment.

**Table A2: Comparison of behavior in dropout and finished groups**

|  | (1) | (2) | (3) | (4) |
|---|---|---|---|---|
| Dependent variable | Effort | Number of neighbors initiated | Number of neighbors | Per period payoffs |
| Dropout group (1=Yes, 0=No) | 0.318 | 0.026 | 0.104 | -7.054 |
|  | (0.305) | (0.167) | (0.181) | (6.675) |
| Period | -0.014* | 0.018*** | 0.022*** | 0.782*** |
|  | (0.008) | (0.003) | (0.003) | (0.147) |
| *N5_HighCost=1 (N5_LowCost=0)* | -0.051 | -0.162 | -0.127 | -3.945 |
|  | (0.294) | (0.144) | (0.136) | (2.594) |
| *N9_LowCost1=1 (N5_LowCost=0)* | 0.334 | 1.155*** | 2.073*** | -0.789 |
|  | (0.294) | (0.187) | (0.211) | (4.258) |
| *N9_HighCost=1 (N5_LowCost=0)* | 0.191 | 0.767*** | 1.594*** | -5.068 |
|  | (0.272) | (0.231) | (0.269) | (3.934) |
| *N9_LowCost2=1 (N5_LowCost=0)* | 2.478*** | 1.770*** | 2.725*** | 70.982*** |
|  | (0.493) | (0.237) | (0.234) | (12.622) |
| Constant | 5.123*** | 2.023*** | 2.952*** | 5.824* |
|  | (0.218) | (0.124) | (0.119) | (3.119) |
| Observations | 15,197 | 15,197 | 15,197 | 15,197 |

*Note:* Multi-level mixed-effects regressions with standard errors clustered at the group level (reported in parenthesis). Stars indicate significance levels: *** 1%, ** 5%, *10%. Sample: all treatments.

### 2.2 The influence of time to form a group

In this section, we analyze if the time spent until forming a group influences the behavior in terms effort and number of neighbors chosen. Since in the experiment, we form groups by the arrival time, which may not be completely random, it might be the case that faster groups exhibit different behavior from slower groups due to their different composition of members. For each group, we calculate the average arrival time in seconds to the group formation page of their members, which becomes a group level variable. The arrival time to the group formation page includes the time spent reading the instructions, answering the control questions and going through the practice periods. The average is 909.595 seconds, with standard deviation of 326.061 across all groups. Then, we calculate the median arrival time across groups by treatment and introduce a dummy variable which takes the value 1 if the group's arrival time is above the median, and 0 if it is below. We run multi-level random effects regressions of individual effort and number of neighbors initiated to study whether being in a slow versus fast-formed group has any impact on the behavior. The results are shown in Table A3. We find that the mentioned dummy variable capturing the relative time to form the group has no significant impact on the individuals' decisions. We also compare the fast and slow groups using non-parametric tests comparing the average effort and number of neighbors initiated at the group level. However, we do not find any significant difference. Test results are summarized in Table A4. These results indicate that the groups formed faster do not exhibit different behavior from those that are formed slower.

**Table A3: Impact of group formation arrival time on decisions**

|  | (1) | (2) |
|---|---|---|
| Dependent variable | Number of neighbors initiated | Effort |
| Group formation took above median time (1=Yes, 0=No) | 0.220 | 0.227 |
|  | (0.169) | (0.295) |
| Period | 0.017*** | -0.013 |
|  | (0.003) | (0.009) |
| Gender (1=Female, 0=Male) | -0.344** | -0.133 |
|  | (0.150) | (0.180) |
| Age | -0.008 | -0.003 |
|  | (0.009) | (0.012) |
| Social preferences | -0.001 | 0.004* |
|  | (0.002) | (0.002) |
| Cognitive Reflection test | 0.299*** | -0.027 |
|  | (0.058) | (0.083) |
| Risk Aversion | 0.001 | 0.004 |
|  | (0.002) | (0.003) |
| Student status (1=Yes, 2=No) | -0.164 | 0.030 |
|  | (0.149) | (0.192) |
| Education in years | -0.050 | 0.030 |
|  | (0.051) | (0.051) |
| ***Treatment dummies*** |  |  |
| N5_HighCost=1 (N5_LowCost=0) | -0.206 | -0.062 |
|  | (0.144) | (0.329) |
| N9_LowCost1=1 (N5_LowCost=0) | 1.207*** | 0.357 |
|  | (0.177) | (0.346) |
| N9_HighCost=1 (N5_LowCost=0) | 0.604** | 0.176 |
|  | (0.288) | (0.273) |
| N9_LowCost2=1 (N5_LowCost=0) | 1.808*** | 2.573*** |
|  | (0.254) | (0.634) |
| Constant | 2.851*** | 4.052*** |
|  | (0.836) | (1.137) |
| Observations | 12,480 | 12,480 |

*Note:* Multi-level mixed-effects regressions with standard errors clustered at the group level (reported in parenthesis). Stars indicate significance levels: *** 1%, ** 5%, *10%. Sample: all treatments.

**Table A4: Comparison of decisions in slow and fast-formed groups by non-parametric tests**

| Treatments | N5_LowCost | N5_HighCost | N9_LowCost1 | N9_HighCost | N9_LowCost2 | All treatments |
|---|---|---|---|---|---|---|
| Slow groups average effort (std. dev) | 5.236 (0.836) | 4.637 (1.106) | 5.387 (0.616) | 5.564 (0.427) | 7.213 (1.987) | 5.515 (1.337) |
| Fast groups average effort (std. dev) | 4.608 (0.524) | 4.973 (1.097) | 5.171 (1.321) | 4.555 (0.566) | 7.702 (2.096) | 5.283 (1.555) |
| Comparison MW-test z-score (p-value) | -1.620 (0.121) | 0.694 (0.536) | -0.940 (0.421) | -2.611*** (0.008) | 0.522 (0.691) | -1.235 (0.221) |
| Slow groups average nr. neighbors initiated (std. dev) | 2.13 (0.549) | 2.000 (0.354) | 3.922 (0.280) | 3.190 (0.611) | 3.670 (0.786) | 2.958 (.0862) |
| Fast groups average nr. neighbors initiated (std. dev) | 2.554 (0.364) | 2.271 (0.346) | 3.171 (0.581) | 2.751 (1.018) | 4.613 (0.639) | 2.835 (1.049) |
| Comparison MW-test z-score (p-value) | -1.620 (0.121) | 1.332 (0.199) | -2.193** (0.032) | -0.940 (0.421) | 1.984* (0.056) | -0.806 (0.425) |

*Note:* We take averages of outcomes over all periods for each group to form independent observations and compare slow and fast formed groups by Mann-Whitney tests. Stars attached to the average numbers indicate significance levels of these tests: *** 1%, ** 5%, *10%.

# 3. Further analyses of linking decisions

## 3.1 Profitability of linking decisions and reciprocated links

In this section, we present further results on the linking decisions. First, we study whether the missing links could indeed be profitably added to the network. When deciding about linking, individuals need to compare the costs and benefits of establishing a link. The benefits of establishing a link between individual $i$ and $j$ are $\lambda x_i x_j$, which depend on the effort choices of both individuals, this is compared to the linking cost $\kappa$. It is not beneficial to establish a link if the opponent's effort level is relatively low. It might be the case that there is large heterogeneity in the effort level within a group in the experimental data and the missing links are those that would link the individual to a low effort provider and hence are unprofitable.

In Table A5, we investigate whether this is the case. We report the average number of links of an individual that could be profitably added given the actual effort choices of opponents in the data. We also report this number as a percentage of the total number of links an individual misses (if any). We consider the last 10 periods of the experiment. We can see that the vast majority, at least 94%, of the missing links could be profitably added when the linking costs are low (as in N5_LowCost, N9_LowCost1, and N9_LowCost2). Under high linking cost, however, we obtain lower figures. 72.6% of the missing links could be profitable added in N5_HighCost and the same number is 76.1% in N9_HighCost. This means that about 25% of the links are unprofitable to add in these treatments, which explains some of the missing links.

While the missing of profitable links seems to be the main reason for welfare loss in the experiment, we identify further welfare-decreasing effects that are connected to excessive linking. In Table A5 we report the number and percentage of links that could be profitably deleted given the effort choices of opponents. The results indicate that in N5_HighCost, 18.6% of the existing links could be profitably deleted; the same number is 16% in N9_HighCost. In contrast, in the treatments with low linking costs, it is not profitable to delete almost any link.

Another source of welfare-loss is the existence of links that are sponsored by both individuals involved. Note that in the Nash equilibrium each link is initiated and sponsored only by one individual involved, since initiating the link by the second person as well only leads to extra costs but no benefits. In practice, however, it is difficult to coordinate on which party involved should sponsor the link. This leads to some links being sponsored by both individuals which reduces the welfare relative to the equilibrium payoffs. In the last column of Table A5, we report the fraction of reciprocated links that are initiated and sponsored by both individuals. We can see that a significant fraction, between 24-45%, of the links fall into this category, which leads to the loss of payoffs for the individuals.

**Table A5: Summary of profitable changes of linking decisions in the last 10 periods**

|  | Avg. #profitable missing links | #profitable missing links/ #missing links | Avg. #unprofitable exiting links | #unprofitable existing links/ #existing links | % of reciprocated links |
|---|---|---|---|---|---|
| N5_LowCost | 0.651 | 0.987 | 0.051 | 0.014 | 0.444 |
| N5_HighCost | 0.573 | 0.726 | 0.571 | 0.186 | 0.355 |
| N9_LowCost1 | 2.315 | 0.949 | 0.115 | 0.024 | 0.301 |
| N9_LowCost2 | 1.667 | 0.988 | 0.022 | 0.005 | 0.355 |
| N9_HighCost | 2.389 | 0.761 | 0.698 | 0.16 | 0.243 |

### 3.2 Analysis of linking in the additional treatment (section 5.2)

In Table A6, we run logistic regressions, similar to the ones presented in Table 8 in the main text, for the additional treatment explained in section 5.2. The results show that individuals link to top effort providers and do not link to relative low effort providers even if the benefits of each link is made explicit to them in the additional treatment. Their behavior is thus the same independently of whether the link benefit information is provided or not.

| Table A6: Relative position in the group and linking | | |
|---|---|---|
| Dependent variable: Link initiated=1 (no=0) | (1) | (3) |
| Reference category | Median | Average |
| Link initiated in t-1 (1=yes, 0=no) | 1.105** | 1.124*** |
|  | (0.046) | (0.048) |
| Opponent's effort in t-1 | 1.021 | 1.033** |
|  | (0.016) | (0.016) |
| Above group median effort in t-1 | 1.923*** |  |
|  | (0.225) |  |
| Below group median effort in t-1 | 0.787*** |  |
|  | (0.071) |  |
| Below average effort in t-1 |  | 0.525*** |
|  |  | (0.066) |
| Constant | 0.766 | 1.134 |
|  | (0.170) | (0.237) |
| Observations | 19,440 | 19,440 |

*Note:* Individual random-effects regressions at the link level with standard errors clustered at the individual level (reported in parenthesis). Odds ratios are reported. Dependent variable: initiating a link (1 if the individual initiated a link to a given opponent, 0 otherwise). Stars indicate significance levels: *** 1%, ** 5%, *10%. Sample: Additional treatment presented in section 5.2 in the main text.

### 3.3 Missing links due to strategic considerations

In this section, we study the possibility that individuals do not link to each other due to coordination failures. Since in the experiment the mutually beneficial links are established unilaterally and only one person involved in the link pays the linking costs, a coordination problem, similar to the battle of sexes, arises, whereby each of the two equilibria is preferred by one of the persons who does not initiate the costly link. The literature on coordination games (see e.g., Arifovic and Ledyard, 2018; He and Wu, 2020) shows that in some cases individuals are able to coordinate on an alternation strategy in which in one period they play the equilibrium preferred by one individual, and in the next period they play the equilibrium preferred by the other. In our game, it might be the case that individuals involved in a link try to coordinate on such an alternation strategy but fail to do so which leads to none of the partners initiating the link. We study this possibility by running another logistic regression with the binary dependent variable being whether individual $i$ initiates a link to individual $j$ in period $t$. The main independent variables of interests are the lagged dependent variable, a dummy variable that captures whether individual $j$ initiated a link to individual $i$ in the previous period $t-1$, as well as, the interaction term of these two.[26] We also control for the effort levels of the two persons in $t-1$ and the treatment dummy variables. The regression results are shown in Table A7.

---

[26] We include the interaction term because it changes the interpretation of the coefficients of the two dummy variables, the lagged dependent variable and the lagged linking intention of the partner. The interaction term captures the possibility that both parties involved in the link initiated it in the previous period, while the single dummy variables capture that only one of the partners initiated the link while the other did not.

On the one hand, our regression would capture a successful alternation strategy if the lagged dependent variable had a statistically significant, *lower than 1* odds ratio (if individual $i$ initiated a link in $t-1$, should not do it in $t$) and the lagged linking intention of the partner had a significant, *higher than 1* odds ratio (if individual $j$ initiated a link in $t-1$, individual $i$ should do it in $t$). On the other hand, a failure to apply this strategy in a way that leads to no link being formed would imply that after a link was initiated by any of the partners in the previous period, there will be no link initiated by any of them in the next. This means that the lagged dependent variable and the lagged linking intention of the partner would have significant and *lower than 1* odds ratios. Our results in Table A7 do not confirm any of these possibilities because the lagged dependent variable is significant with a *higher than 1* odds ratio in all our regressions, which is the opposite than what was predicted by these explanations. In addition, the lagged linking intention of the partner is not statistically significant in the regression including all treatments and the regression including the treatments with the larger group size ($N = 9$). These findings mean that individuals keep initiating the same link, without strategically considering the linking costs and whether their partner was willing to initiate the link previously. For the regression including the treatments with the smaller group size ($N = 5$), we find some evidence for strategic behavior in the sense that individual $i$ is less likely to initiate a link after their partner initiated it in the previous period (evidenced by the significant and lower than 1 odds ratio of the dummy capturing whether the partner initiated the link in $t-1$). However, this does not lead to the loss of links as the partner may keep initiating the same link. In sum, we find that strategic considerations to save on the linking costs and the resulting coordination failure cannot explain the missing links in our experiment.

### Table A7: Strategic behavior in linking decisions

| Dependent variable: Link initiated=1 (not=0) | (1) | (2) | (3) |
|---|---|---|---|
| Sample | All treatments | Treatments with N=5 | Treatments with N=9 |
| Link initiated in t-1 (1=yes, 0=no) | 2.894*** | 2.003*** | 3.118*** |
|  | (0.204) | (0.233) | (0.254) |
| Partner initiated a link in t-1 | 1.016 | 0.800** | 1.066 |
|  | (0.049) | (0.073) | (0.059) |
| I linked in t-1 * Partner linked in t-1 | 0.926 | 1.166 | 0.908 |
|  | (0.068) | (0.125) | (0.083) |
| Own effort in t-1 | 1.004 | 1.000 | 1.005 |
|  | (0.006) | (0.010) | (0.006) |
| Partner's effort in t-1 | 1.083*** | 1.070*** | 1.085*** |
|  | (0.009) | (0.015) | (0.010) |
| N5_HighCost | 0.830 | 0.817 |  |
|  | (0.139) | (0.142) |  |
| N9_LowCost1 | 0.532*** |  |  |
|  | (0.083) |  |  |
| N9_HighCost | 0.402*** |  | 0.759* |
|  | (0.066) |  | (0.109) |
| N9_LowCost2 | 0.651** |  | 1.209 |
|  | (0.122) |  | (0.205) |
| Constant | 0.598*** | 0.847 | 0.300*** |

|  | (0.084) | (0.144) | (0.036) |
|---|---|---|---|
| Observations | 84,948 | 17,996 | 66,952 |

*Note:* Individual random-effects regressions at the link level with standard errors clustered at the individual level (reported in parenthesis). Odds ratios are reported. Dependent variable: initiating a link (1 if the individual initiated a link to a given opponent, 0 otherwise). Stars indicate significance levels: *** 1%, ** 5%, *10%. Sample: all treatments in column 1, treatments with $N = 5$ in column 2, treatments with $N = 9$ in column 3.

### 3.4 Individuals punish others by not linking to them

Finally, we consider if the missing links can be explained by punishment. Individuals may not link to others whom they think unfairly treated them in the previous round. The lack of links could thus be a sort of retaliation. Behavior that may trigger such retaliation can be either lowered effort or a withdrawn link by the opponent to the individual. We test these ideas by regression analysis where the outcome is a dummy variable capturing if individual $i$ has initiated a link to a given opponent. The explanatory variables are dummy variable capturing whether the opponent lowered effort or withdrew a link to individual $i$ in the previous period. The regression results are shown in Table A8. We find that none of the explanatory variables of interests is significant in the regression. Punishment thus cannot explain the lack of links formed.

### Table A8: Punishment and linking decisions

| Dependent variable: Link initiated=1 (not=0) | (1) | (2) |
|---|---|---|
| Link initiated in t-1 (1=yes, 0=no) | 2.804*** | 2.801*** |
|  | (0.183) | (0.182) |
| Own effort in t-1 | 1.004 | 1.003 |
|  | (0.005) | (0.006) |
| Partner's effort in t-1 | 1.081*** | 1.083*** |
|  | (0.009) | (0.009) |
| Partner initiated a link in t-1 | 0.981 | 0.987 |
|  | (0.033) | (0.035) |
| Partner lowered effort in t-1 | 0.977 |  |
|  | (0.023) |  |
| Partner withdrew link in t-1 |  | 1.066 |
|  |  | (0.060) |
| Lambda | 39.240*** | 37.196*** |
|  | (27.733) | (26.295) |
| Linking costs | 0.976 | 0.975 |
|  | (0.044) | (0.044) |
| Constant | 0.120*** | 0.120*** |
|  | (0.029) | (0.028) |
| Observations | 84,948 | 84,948 |

*Note:* Individual random-effects regressions at the link level with standard errors clustered at the individual level (reported in parenthesis). Odds ratios are reported. Dependent variable: initiating a link (1 if the individual initiated a link to a given opponent, 0 otherwise). Stars indicate significance levels: *** 1%, ** 5%, *10%. Sample: all treatments.

### 3.5 The impact of accumulated payoffs

In this section, we study whether satisficing can explain some of the individual behavior in the late periods of the game. In particular, individuals may stop expanding their number of neighbors because they have already accumulated sufficient amount of payoffs, forming more links and further increasing their payoffs may thus be seen as unnecessary. If this is true, we should observe that those individuals who accumulated more payoffs up to period 20, will initiate less links in the last 10 periods. To study this effect, we regress the number of neighbors initiated by the individual in the last 10 periods of the game on the payoffs accumulated up to period 20, controlling for the number of neighbors in period 20, individual characteristics and the period number when the decision was submitted. If satisficing is at play, the coefficient of accumulated payoffs should be negative as individuals with more payoffs will initiate less links. The regression results in columns (1)-(3) of Table A5 show that the accumulated payoffs has significant positive or insignificant coefficients, indicating that those accumulated higher payoffs up to period 20, do not initiate less links. Satisficing thus cannot explain the linking behavior in the last periods of the game. Similarly, we regress the effort choices in the last 10 periods on the accumulated payoffs up to period 20, controlling for the effort choice in period 20, individual characteristics and the period number when the decision was submitted. We find that accumulated payoffs has an insignificant coefficient in all regressions reported in columns (4)-(6) in Table A9. Accumulated payoffs thus do not influence the effort choice.

### 3.6 Impact of individual characteristics on linking

While the impact of own degree on the linking decision enlarges the differences between group members, a relevant question is what factors explain why these differences emerge at the first place. To answer this question, we study the impact of individual characteristics on the number of links initiated by the individual using the data from the post experimental survey. We run a multi-level random effects regression of the number of links initiated where the independent variables are the individuals' gender, age, social preferences, risk aversion, cognitive-reflection test results, student status and education in years. We pool the data from all treatments to have sufficient variation in these characteristics at the individual level and add treatment dummies to the regression. We cluster standard errors at the group level and run separate regressions for all periods and the last 10 periods only. Table A10 shows the results.

We find two individual characteristics that significantly influence the number of links initiated. The first one is gender which shows that female initiate significantly less links than male. The second is the cognitive-reflection (CR) test result where we obtain that individuals answering more questions correctly on the incentivized test initiate more links. Both of these results are in line with the findings of previous literature. Regarding gender, several recent empirical papers find that female possess a lower number of collaborative relationships than male. Lindenlaub and Prummer (2021) find this gender effect in friendship networks in high schools, in e-mail communication networks in an organization, and considering the scientific collaboration network of computer scientists. Ductor et al. (2021) and Jadidi et al. (2018) find a similar relationship in scientific collaboration networks in other disciplines. Regarding the impact of the cognitive-reflection test results, several studies find that individuals who do better on the test make more rational decisions in games (Branas et al., 2019). More specifically, Branas et al. (2012) and Kiss et al. (2016) find that individuals who score higher on the CR test are more likely to play the dominant strategy, while Carpenter et al. (2013) obtain that they show higher strategic sophistication. In our experiment, creating more of the profitable links is the rational behavior which is positively correlated with the CR test results.

**Table A9: The impact of accumulated payoffs on number of links initiated and effort choices in the last 10 periods of the game**

|  | (1) | (2) | (3) | (4) | (5) | (6) |
|---|---|---|---|---|---|---|
| Dependent variable | Number of neighbors initiated | Number of neighbors initiated | Number of neighbors initiated | Effort | Effort | Effort |
| Sample | All treatments | Treatments with N=5 | Treatments with N=9 | All treatments | Treatments with N=5 | Treatments with N=9 |
| Cumulative payoffs up t=20 | 0.0004*** | 0.0003 | 0.0004*** | -0.0005 | -0.0004* | -0.0004 |
|  | (0.0001) | (0.0002) | (0.0001) | (0.0003) | (0.0002) | (0.0003) |
| # neighbors in t=20 | 0.677*** | 0.469*** | 0.719*** |  |  |  |
|  | (0.059) | (0.068) | (0.069) |  |  |  |
| Effort in t=20 |  |  |  | 0.445*** | 0.201*** | 0.553*** |
|  |  |  |  | (0.078) | (0.040) | (0.090) |
| Period | 0.008 | -0.001 | 0.013 | -0.015 | -0.026 | -0.008 |
|  | (0.007) | (0.007) | (0.011) | (0.013) | (0.017) | (0.018) |
| Gender (1=Female, 0=Male) | -0.158 | -0.172 | -0.135 | -0.050 | -0.193 | 0.001 |
|  | (0.129) | (0.140) | (0.201) | (0.131) | (0.156) | (0.189) |
| Age | -0.009 | -0.003 | -0.012 | -0.001 | -0.003 | -0.003 |
|  | (0.009) | (0.013) | (0.011) | (0.008) | (0.010) | (0.011) |
| Social preferences | 0.001 | -0.002 | 0.003 | 0.000 | -0.000 | 0.001 |
|  | (0.002) | (0.003) | (0.003) | (0.002) | (0.002) | (0.003) |
| Cognitive Reflection test | 0.157*** | 0.061 | 0.218*** | -0.004 | -0.020 | -0.008 |
|  | (0.041) | (0.061) | (0.056) | (0.042) | (0.066) | (0.056) |
| Risk Aversion | 0.000 | 0.002 | -0.001 | -0.001 | -0.001 | -0.000 |
|  | (0.002) | (0.002) | (0.003) | (0.003) | (0.003) | (0.003) |
| Student status (1=Yes, 2=No) | -0.066 | 0.146 | -0.167 | 0.055 | -0.001 | 0.054 |
|  | (0.133) | (0.192) | (0.176) | (0.147) | (0.127) | (0.213) |
| Education in years | 0.008 | -0.059* | 0.053 | -0.024 | -0.034 | -0.011 |
|  | (0.043) | (0.033) | (0.064) | (0.041) | (0.041) | (0.054) |
| ***Treatment dummies*** |  |  |  |  |  |  |
| N5_HighCost=1 | -0.178 | -0.203 |  | 0.029 | 0.071 |  |
|  | (0.143) | (0.138) |  | (0.295) | (0.325) |  |
| N9_LowCost1=1 | -0.546*** |  |  | 0.524** |  |  |
|  | (0.195) |  |  | (0.238) |  |  |
| N9_HighCost=1 | -0.744*** |  | -0.180 | 0.132 |  | -0.330 |
|  | (0.202) |  | (0.135) | (0.263) |  | (0.236) |
| N9_LowCost2=1 | -0.791*** |  | -0.245 | 2.952*** |  | 2.118*** |
|  | (0.248) |  | (0.210) | (0.574) |  | (0.499) |
| Constant | -0.272 | 1.948** | -2.107 | 3.350*** | 5.161*** | 2.856** |
|  | (0.880) | (0.832) | (1.329) | (0.866) | (0.840) | (1.237) |
| Observations | 4,160 | 1,460 | 2,700 | 4,160 | 1,460 | 2,700 |

*Note:* Multi-level mixed-effects regressions with standard errors clustered at the group level (reported in parenthesis). Dependent variable: number of links initiated. Stars indicate significance levels: *** 1%, ** 5%, *10%.

**Table A10: Regression of the number of links initiated on individual characteristics**

| Dependent variable: Number of links initiated | (1) All periods | (2) Last 10 periods |
|---|---|---|
| Gender (1=Female, 0=Male) | -0.325** | -0.322** |
|  | (0.149) | (0.160) |
| Age | -0.008 | -0.007 |
|  | (0.009) | (0.010) |
| Social preferences | -0.001 | -0.001 |
|  | (0.002) | (0.003) |
| Cognitive Reflection test | 0.290*** | 0.276*** |
|  | (0.058) | (0.058) |
| Risk Aversion | 0.001 | 0.001 |
|  | (0.002) | (0.003) |
| Student status (1=Yes, 2=No) | -0.164 | -0.082 |
|  | (0.149) | (0.174) |
| Education in years | -0.051 | -0.033 |
|  | (0.051) | (0.055) |
| *Treatment dummies* | | |
| N5_HighCost=1 (N5_LowCost=0) | -0.207 | -0.261 |
|  | (0.154) | (0.189) |
| N9_LowCost1=1 (N5_LowCost=0) | 1.217*** | 1.237*** |
|  | (0.207) | (0.248) |
| N9_HighCost=1 (N5_LowCost=0) | 0.614** | 0.542* |
|  | (0.297) | (0.290) |
| N9_LowCost2=1 (N5_LowCost=0) | 1.817*** | 1.877*** |
|  | (0.245) | (0.244) |
| Constant | 3.270*** | 2.992*** |
|  | (0.817) | (0.907) |
| Observations | 12,480 | 4,160 |

*Note:* Multi-level mixed-effects regressions with standard errors clustered at the group level (reported in parenthesis). Dependent variable: number of links initiated. Stars indicate significance levels: *** 1%, ** 5%, *10%. Column (1) reports the regression with data from all periods, column (2) with data from the last 10 periods. Data from all treatments are pooled and treatment dummies are added to the regression, with the omitted category being N5_LowCost.

# 4. Additional figures

**Figure A1: An example of the decision screen**

**Decision in Round 1**

Time left to complete this page: **0:57**

Your ID number is ID9.

Please, choose whom you would like to link to:
- ☐ Id1
- ☐ Id2
- ☐ Id3
- ☐ Id4
- ☐ Id5
- ☐ Id6
- ☐ Id7
- ☐ Id8

Please, choose an activity level:

[              ]

[Next]

**Figure A2: An example of the feedback screen at the end of a period**

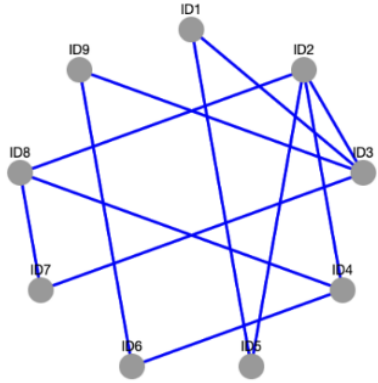

# Figure A3: Feedback screen in the additional treatment with information on link benefits (section 5.2)

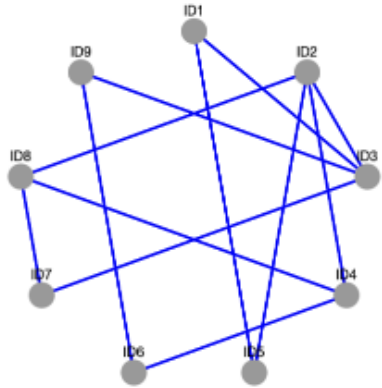

**Figure A4: Relative connectivity compared to the complete network and average degree over time and by treatments**

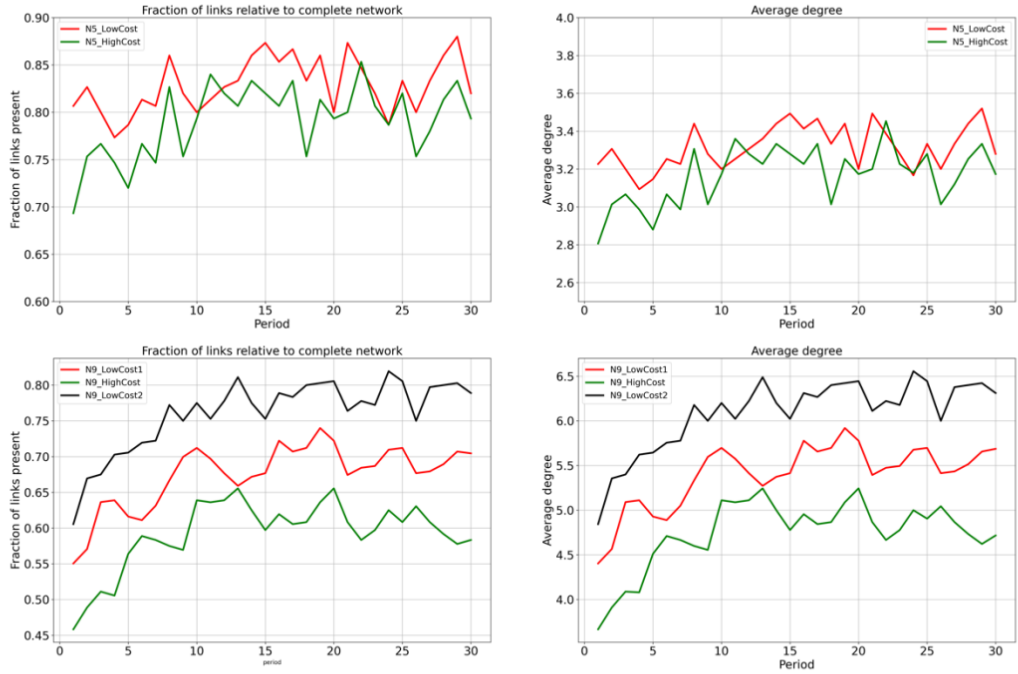

**Figure A5: Average effort and relative efficiency compared to the complete network equilibrium level over time and by treatment**

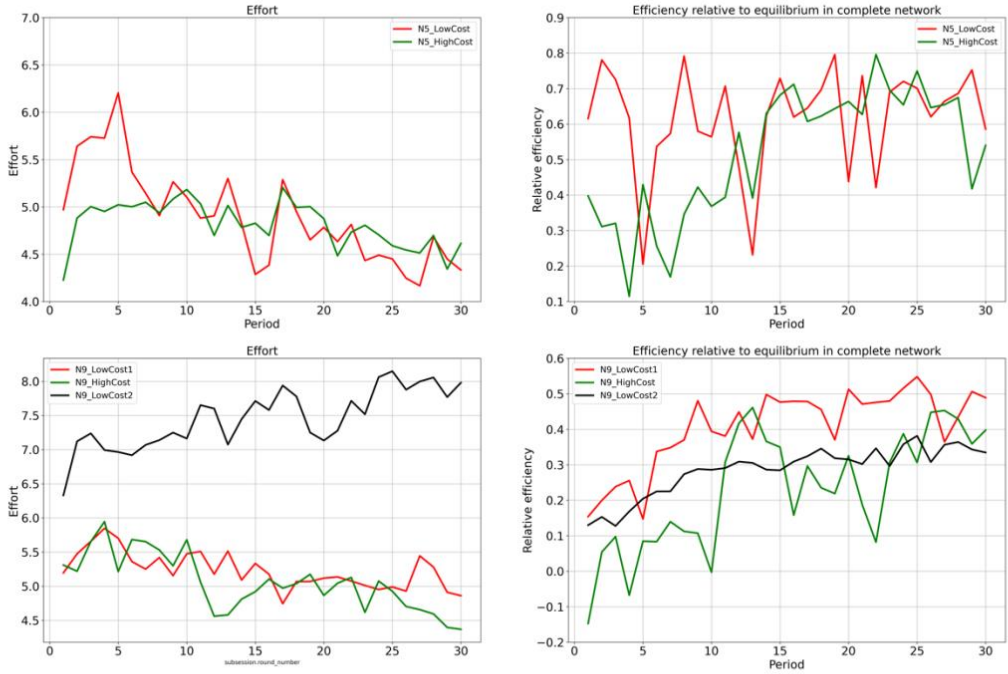

**Figure A6: Effort dynamics and fitted values from the myopic best-response model presented in section 5.1**

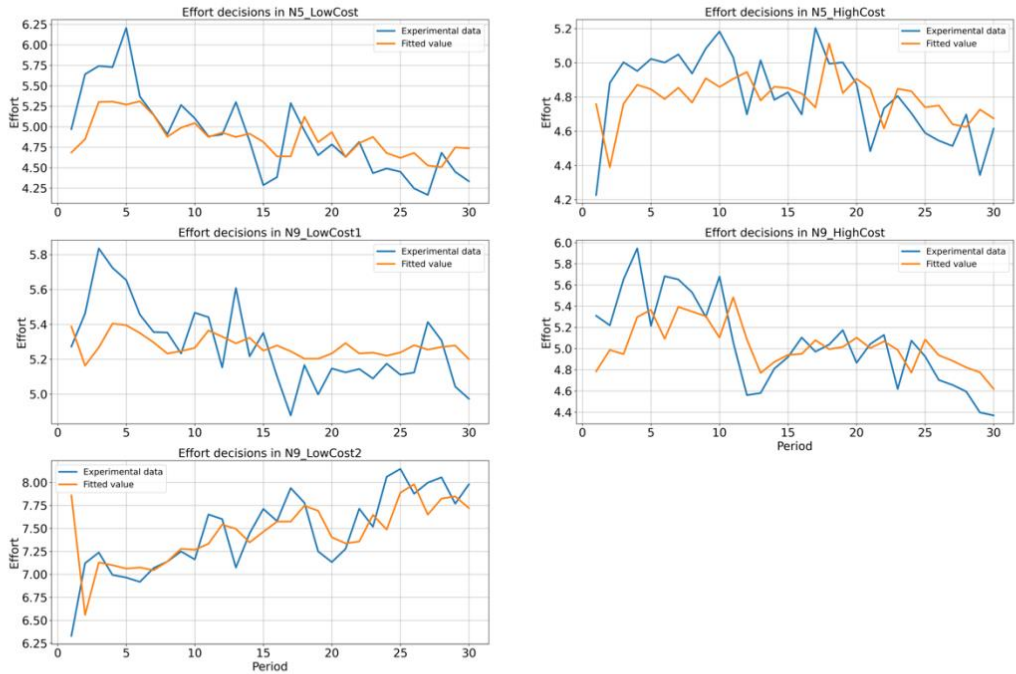

**Figure A7: Average benefit per missing link and per period payoffs over time**

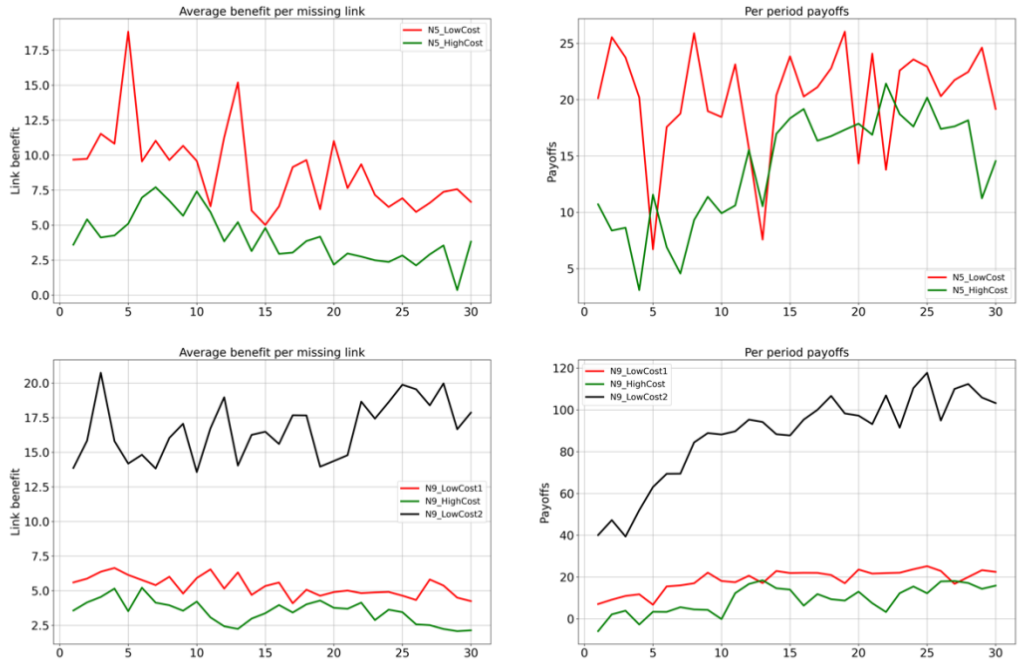

**Figure A8: Network structures formed in the last 5 periods of the experiment in *N5_LowCost***

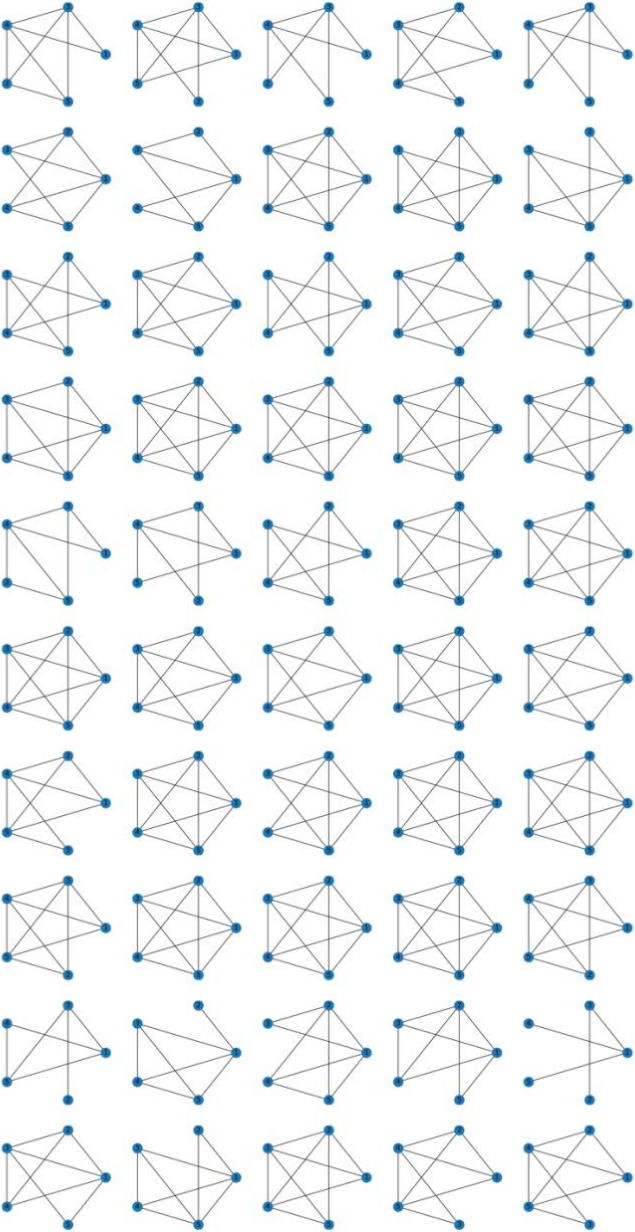

**Figure A9: Network structures formed in the last 5 periods of the experiment in *N5_HighCost***

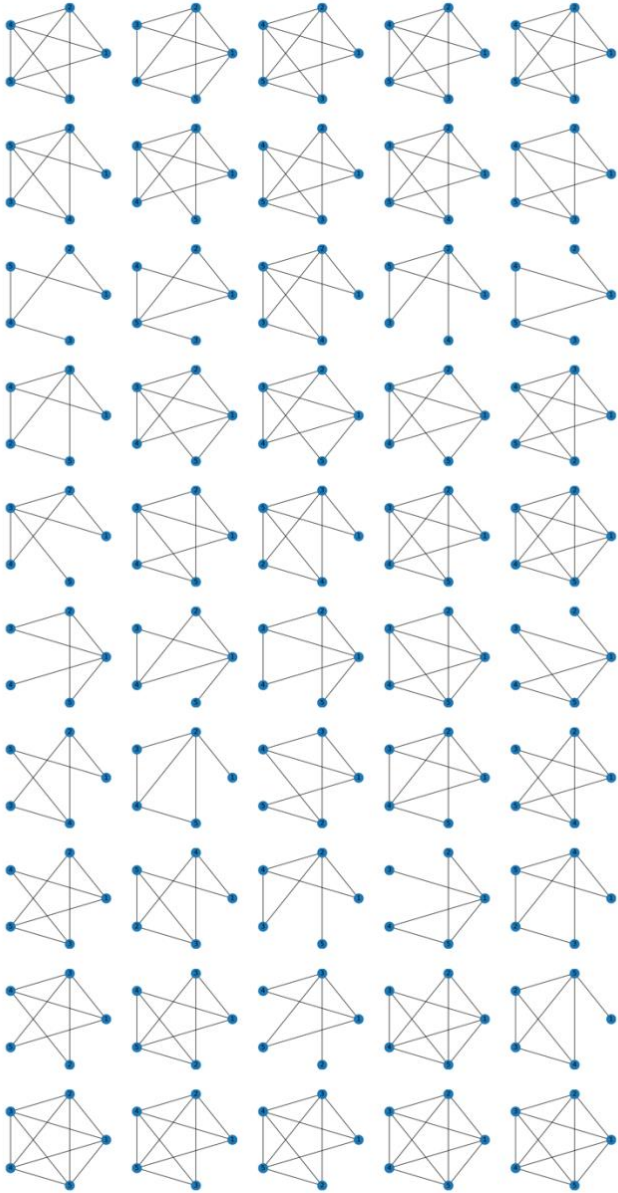

**Figure A10: Network structures formed in the last 5 periods of the experiment in *N9_LowCost1***

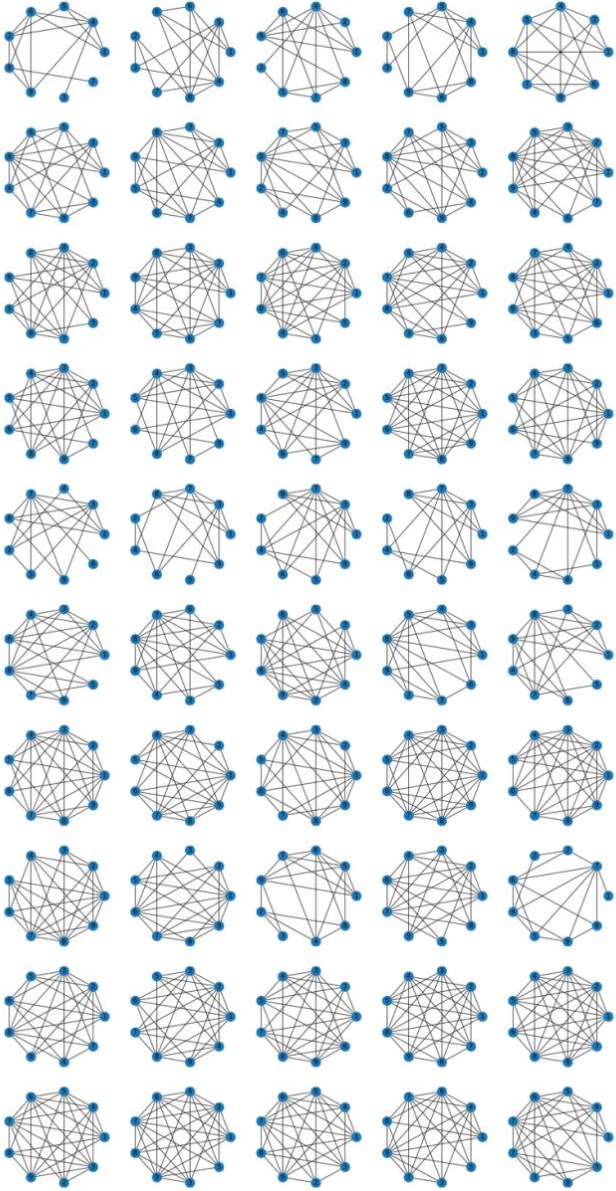

**Figure A11: Network structures formed in the last 5 periods of the experiment in** *N9_HighCost2*

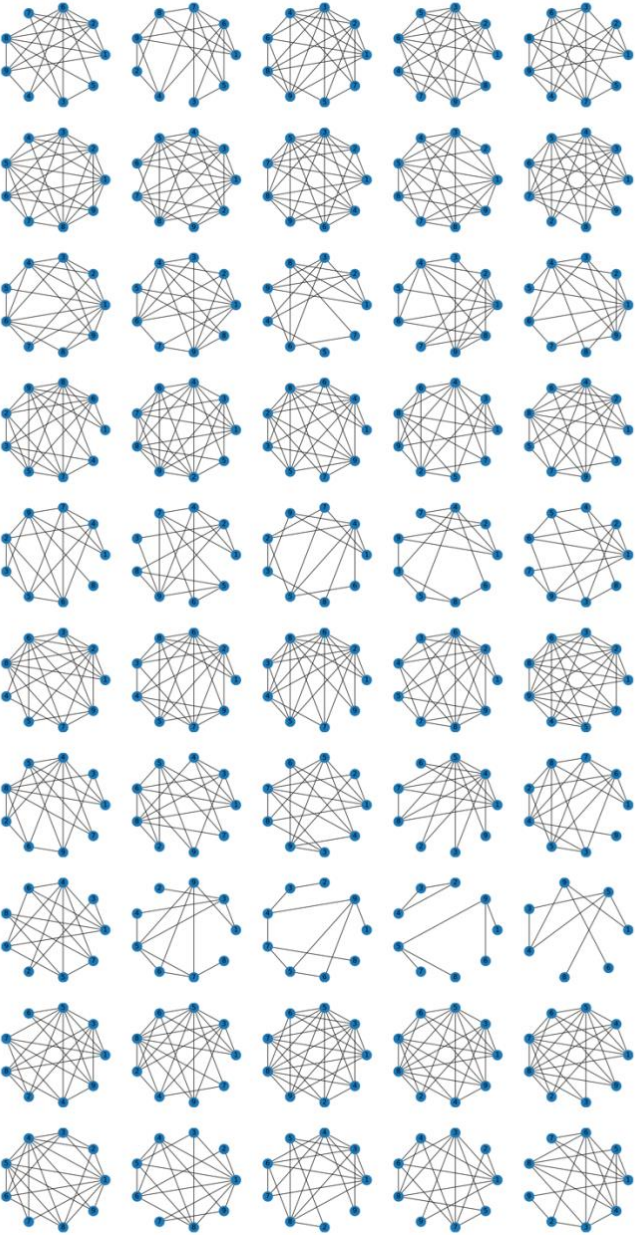

**Figure A12: Network structures formed in the last 5 periods of the experiment in *N9_LowCost2***

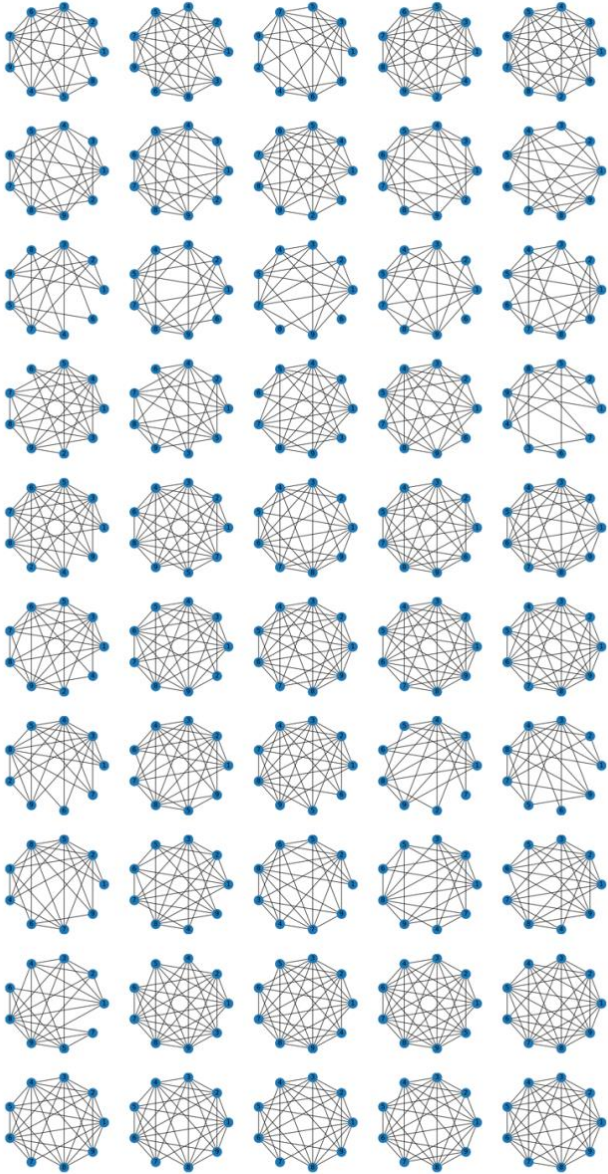